\documentclass[10pt,sigplan,letterpaper,twocolumn]{article}
\usepackage[10pt,nocopyright]{sigmin}

\usepackage{listings}
\usepackage{color}
\usepackage{dingbat}
\usepackage[short]{optidef} 

\usepackage{pgfplots}
\pgfplotsset{compat=1.17}

\usepackage{epsfig,makecell}                                      
\usepackage[hyphens]{url}                                         
\usepackage[breaklinks,colorlinks]{hyperref}                      
\hypersetup{citecolor=blue,linkcolor=blue}                        
\usepackage{amsmath,amsopn,amssymb,amsthm}                        
\usepackage{thmtools}
\usepackage{thmbox}
\usepackage[toc,page]{appendix}
\usepackage{subcaption}                                           
\usepackage{endnotes,microtype,xspace,fancyvrb,multirow} 
\usepackage{epigraph} 
\usepackage{stfloats}

\usepackage{booktabs}      
\usepackage{tabularx}
\usepackage{array,underscore,relsize}                             
\usepackage[T1]{fontenc}                                          
\usepackage{times}                                                
\usepackage{fancyhdr,lastpage}                                    
\usepackage{enumitem}                                             
\usepackage[labelfont=bf,font=normal,skip=1pt]{caption}           
\usepackage[export]{adjustbox}                                    
\usepackage{breakurl}                                             
\usepackage{pifont}                                               
\usepackage{tablefootnote}                                        
\usepackage{bbding}                                        
\usepackage[linesnumbered,vlined,boxed,commentsnumbered,ruled]{algorithm2e}
\usepackage{float}
\DontPrintSemicolon
\newcommand{\ra}[1]{\renewcommand{\arraystretch}{#1}}
\usepackage{amssymb}
\usepackage[normalem]{ulem}
\usepackage[hang,flushmargin]{footmisc}
\usepackage{textcomp}
\usepackage{graphicx}
\usepackage{authblk}

\usepackage[compact]{titlesec}
\titleformat*{\section}{\large\bfseries}
\titleformat*{\subsection}{\normalsize\bfseries}
\titleformat*{\subsubsection}{\normalsize\bfseries}
\titlespacing{\section}{0pt}{3ex}{1ex}
\titlespacing{\subsection}{0pt}{2ex}{1ex}

\hypersetup{
  colorlinks,
  linkcolor={red!50!black},
  citecolor={blue!50!black},
  urlcolor={blue!80!black}
}

\newcommand{\figsuffix}{.png} 

\newcommand{\kvcache}{\textsc{KV\$}}
\newcommand{\policy}{WA}

\newcommand{\company}{\textsc{Aliyun}}
\newcommand{\companyfull}{\textsc{Aliyun Tongyi}}

\newcommand{\fig}[1]{Figure{~\ref{#1}}}

\newcommand{\alg}[1]{Algorithm{~\ref{#1}}}
\newcommand{\tbl}[1]{Table{~\ref{#1}}}

\newcommand{\cmark}{\ding{51}}%
\newcommand{\xmark}{\ding{55}}%
\newcommand{\one}{\texttt{\uppercase\expandafter{\romannumeral1}}}
\newcommand{\two}{\texttt{\uppercase\expandafter{\romannumeral2}}}

\newcommand{\pcomment}[2]{{\bf[\textcolor{red}{#1}: \textcolor{blue}{#2}]}}

\newcommand{\SSJ}[1]{{\pcomment{SSJ}{#1}}}

\newcommand{\nospacestitle}[1]{\noindent{\bf #1}}
\newcommand{\stitle}[1]{\vspace{1.1ex}\noindent{\bf #1}}

\usepackage{tikz}
\usepackage{colortbl}
\usetikzlibrary{tikzmark,calc}

\usepackage{balance}
\newcolumntype{P}[1]{>{\centering\arraybackslash}p{#1}}

\definecolor{appcolor}{RGB}{191,255,255}

\SetKwInput{KwInput}{Input}
\SetKwInput{KwOutput}{Output}
\SetKwComment{Comment}{$\triangleright$\ }{}

\makeatletter
\renewcommand\AB@affilsepx{ \quad\protect\Affilfont \, } 
\makeatother

\begin{document}

\title{\Large \bf{KVCache Cache in the Wild: Characterizing and Optimizing KVCache Cache \\ at a Large Cloud Provider}}

\setlength{\affilsep}{0.3em}
\author[1]{Jiahao Wang$^\dagger$}
\author[1]{Jinbo Han$^\dagger$}
\author[1]{Xingda Wei\,{\Envelope}}
\author[2]{Sijie Shen}
\author[1]{Dingyan Zhang}
\author[2]{Chenguang Fang}
\author[1]{Rong Chen}
\author[2]{Wenyuan Yu}
\author[1]{Haibo Chen}

\affil[1]{Institute of Parallel and Distributed Systems, Shanghai Jiao Tong University}
\affil[2]{Alibaba Group\vspace{-1.mm}}

\date{}
\maketitle

\def\thefootnote{$\dagger$}\footnotetext{The two authors contributed equally to this work.}
\def\thefootnote{\Envelope}\footnotetext{Xingda Wei is the corresponding author (\url{wxdwfc@sjtu.edu.cn}).}

\renewcommand{\thefootnote}{[\arabic{footnote}]}

\frenchspacing

\begin{abstract}
    \noindent
    Serving large language models (LLMs) is important for cloud providers,
    and caching intermediate results ({\kvcache}) after processing each request
    substantially improves serving throughput and latency.
    However, there is limited understanding of how LLM serving benefits
    from {\kvcache} caching,
    where system design decisions like cache eviction policies
    are highly workload-dependent.

    In this paper, we present the first systematic characterization of
    the {\kvcache} workload patterns from one of the leading LLM service providers.
    We draw observations that were not covered
    by previous studies focusing on synthetic workloads,
    including: {\kvcache} reuses are skewed across requests,
    where reuses between single-turn requests are equally important as multi-turn requests;
    the reuse time and probability are diverse considering all requests,
    but for a specific request category, the pattern tends to be predictable;
    and the overall cache size required for an ideal cache hit ratio is moderate.
    Based on the characterization,
    we further propose a workload-aware cache eviction policy
    that improves the serving performance under real-world traces,
    especially with limited cache capacity.
\end{abstract}

\section{Introduction}
\label{sec:intro}

\noindent
Large language models (LLMs) have become an important cloud service
thanks to their emerging capabilities in various tasks.
Both customer-facing services (to-C) like conversation-based chatbots~\cite{chatgpt}
and business services (to-B) through LLM API calls have been widely deployed on the cloud~\cite{openaiapi,genimiapi,claudeapi}.
As online services, serving LLMs with low latency and high throughput is critical,
but achieving this in a cost-efficient way is challenging
due to the huge computational requirements to serve each LLM request.

Caching intermediate results of served requests ({\kvcache}) has become a common methodology
to improve serving performance~\cite{openai-cache,claude-cache}:
when two requests share the same input prefix,
their {\kvcache} are identical;
so if we cache the {\kvcache} of the first ({\kvcache} cache),
we don't need to compute the {\kvcache} for the next.
Hence, the serving latency is reduced with the throughput improved
thanks to the reduced computation under hits.

Like traditional data caching systems~\cite{yang2020twemcache,atikoglu2012workload,180324},
the effectiveness of {\kvcache} cache system components 
like cache eviction policy should be tailored to the {\kvcache} reuse features of workloads.
However,
LLM serving workloads are complex due to their diverse request types
as well as the unique features of each request type.
For example, a production LLM service handles both to-C workloads (like chatting)
and to-B workloads (like task automation with API calling).
These requests can further be processed either as single-turn or multi-turn interactions (\textsection{\ref{sec:bg}}).
Since the {\kvcache} can be reused across different single-turn requests and multi-turn requests,
both the single-turn request types and the multi-turn patterns
can possibly affect the design choices of {\kvcache} cache.

There remains a significant gap in the understanding of common LLM serving
workloads for {\kvcache} cache due to the lack of in-depth analysis of real-world traces.
First, it is unclear whether production workloads have many reuses and which request
type contributes most to the reuses.
Second, the distributions of the reuse time or reuse probability of {\kvcache} remain unknown,
which is critical to cache policy designs.
Finally, there is a lack of characterization of the lifespan of {\kvcache},
which is essential in helping us estimate the required cache capacity for serving LLMs in the wild.

\begin{table}[!t]
    \centering
    \small{
        \resizebox{1\linewidth}{!}{
            \ra{1.2}
            \begin{tabular}{lrrrrr}
                \toprule
                & \multicolumn{1}{c}{\textbf{Time}} & \multicolumn{1}{c}{\textbf{Type}} & \multicolumn{1}{c}{\textbf{UID}} & \multicolumn{1}{c}{\textbf{Turn}}     & \multicolumn{1}{c}{\textbf{Content}} \\ \hline
\textbf{ShareGPT~\cite{sharegpt}}  &      \xmark                    &   \xmark                    &        \xmark           & \cmark                                  &       \cmark                               \\
\textbf{Mooncake~\cite{mooncake}}  &      \cmark                    &   \xmark                    &        \xmark           & \xmark                                  &       \cmark                               \\
\textbf{Traces in this work}       &      \cmark                    &   \cmark                    &        \cmark           & \cmark                                  &       \cmark                               \\ 
                \bottomrule
\end{tabular}
            
        }
    } \\[8pt]
    \begin{minipage}{1\linewidth}
        \caption{\small{\emph{
        A comparison of currently available LLM serving traces. \textbf{Time} is the invocation time of each request. 
        \textbf{Type} is the request type, e.g., whether it is an API call. 
        \textbf{UID} is the user ID that invokes the request. 
        \textbf{Turn} is the multi-turn information of the request. 
        \textbf{Content} is the (anonymous) input and output tokens of each request.
        }}}
    \label{tab:compare-traces}
    \end{minipage} \\[-15pt]
\end{table}


            

\stitle{Characterizing production traces. \,}
To bridge this gap,
we collected and characterized two representative production serving workloads
from one of the world's largest cloud providers (\company):
a to-C workload where users submit requests to the cloud-hosted LLM service via browser-based
chatbots or mobile applications, and a to-B workload where developers
send OpenAI-compatible requests to the LLM service through API calls in their programs.
We collected as much detailed information as possible within the company's strict privacy policy,
including request submission times, request types (single-turn, multi-turn, API requests, or others),
and anonymized content to analyze caching effectiveness.
To our knowledge, no other available LLM serving traces
match the information depth and scale of our traces (summarized in Table~\ref{tab:compare-traces}).
For example, ShareGPT~\cite{sharegpt} only provides the input and output of each request,
not the submission time of each request,
which is critical for analyzing the real {\kvcache} hit ratio and usage over time.
We believe our collected workloads represent typical LLM serving workloads
and can provide a foundation for future {\kvcache} cache system design.

From the perspective of designing an efficient {\kvcache} cache system---both in performance
and resource usage,
our \emph{key takeaways} are as follows: \\[-15pt]
\begin{enumerate}[leftmargin=*,leftmargin=10pt,itemindent=0pt]
    \item {\kvcache} reuses are common,
    but the reuse ratio is smaller than previously reported numbers
    on synthetic datasets~\cite{cachedattention,DBLP:journals/corr/abs-2403-14401}.
    The reuses follow a skewed distribution:
    10\,\% of {\kvcache} blocks contribute to 77\,\% of the reuses.
    The contribution of {\kvcache} reuses also varies across traces:
    contrary to a straightforward observation that multi-turn requests may dominate most reuses,
    single-turn requests dominate 97\,\%
    {\kvcache} reuses in to-B workloads (\textsection{\ref{sec:workload-patterns}}).  \\[-15pt]

    \item Key workload features related to {\kvcache} designs,
    like reuse time and reuse probability distributions---vary across different LLM request types (API vs. chat)
    and the number of turns in multi-turn requests.
    Despite this diversity, for each specific request category (type plus turn number),
    the reuse time is predictable based on the historical information (\textsection{\ref{sec:motiv-temporal-spatial}}).
    Since the request categories are known by the serving system,
    we can systematically use the estimated probability for improving cache policies. \\[-15pt]

    \item The lifespan of {\kvcache} is ephemeral: the P99 lifespan
    of {\kvcache} in to-B workloads is 97 seconds,
    so a small cache is typically sufficient for the {\kvcache} cache.
    For example, in the to-B workload,
    a {\kvcache} with capacity 2\,$\times$ of the GPU HBM per-GPU
    is sufficient to approach an ideal hit rate under infinite capacity on
    common GQA models~\cite{DBLP:conf/emnlp/AinslieLJZLS23}
    using standard cache eviction policies like LRU (\textsection{\ref{sec:analyze-capacity}}).
\end{enumerate}

Our characterizations above reveal several important design decisions
that current {\kvcache} cache systems may have overlooked.
First, optimizing caching for single-turn requests is equally important as optimizing
for multi-turn requests.
Second, we could improve cache hits
by adopting a workload-aware caching policy
instead of only adopting a workload-agnostic policy like LRU.
Finally, the ephemeral lifespan of {\kvcache} suggests that
a least-frequently-used (LFU) policy is not suitable because a short-lived {\kvcache} 
may have high frequency in the past. 
Meanwhile, 
a small (on-GPU) cache is sufficient for API-dominated to-B workloads,
eliminating the cost and complexity of
deploying and managing a CPU-RDMA-SSD storage hierarchy for {\kvcache} cache.
Besides the main takeaways described above, 
we also discovered several interesting findings, 
such as the {\kvcache} hits across users being extremely low (\textsection{\ref{sec:workload-patterns}}),
despite the possibility of shared prompts via prompt libraries~\cite{prompt-library}. 

\stitle{Improved {\kvcache} cache eviction policy. \,}
Based on our findings,
we propose a workload-aware {\kvcache} eviction policy (\textsection{\ref{sec:design-policy}}) that leverages
the profiled reuse probability distributions of each workload to enhance cache hit ratios
on real-world workloads.
Specifically, we leverage the empirically sampled probability distributions of the reuse probability 
of each workload to decide the priority of each {\kvcache} block, 
with considerations of both the characterized spatial locality and the lifespan of {\kvcache}.
To demonstrate the effectiveness of our improved design,
we have integrated our policy into vLLM~\cite{vllm}---a popular LLM serving system on the cloud,
and our extensive evaluations on production traces show that
the improved design can improve 3.9\,\% cache hits
as well as up to 41.4\,\% mean response time improvements
compared to workload-agnostic policies like LRU and LFU.

\stitle{Discussion. \,}
Before moving on, we note two limitations of our work.
First, our results are based on one week production traces of large-scale LLM serving workloads.
While we believe such workloads (chatbot and API) are representative nowadays,
LLM serving workloads are evolving, with new workloads like reasoning~\cite{openai-o1} emerging.
We leave the characterization of these newer workloads to future work.
Second, our primary focus is on the {\kvcache} cache policy design,
yet we believe the design of other system components for LLM serving 
like the policy of global scheduling 
could also benefit from our findings,
and we leave other explorations as future work.
Nevertheless, we believe our methodology---first characterizing real-world serving features
and then optimizing a specific serving component---generalizes to other LLM serving workloads and systems.

A sample of our analyzed trace is available at \burl{https://github.com/alibaba-edu/qwen-bailian-usagetraces-anon},
and the reference implementation of the eviction policy is available at
\burl{https://github.com/vllm-project/vllm/pull/22236}. 

\section{Background: LLM serving, KVCache, KVCache cache and Multi-turn Serving}
\label{sec:bg}

\begin{figure}[!t]
        \begin{minipage}{1\linewidth}
        \hspace{-1mm}
        \centering    
        \includegraphics[width=.9\columnwidth, trim=0.25cm 12.7cm 30cm 0.25cm, clip]{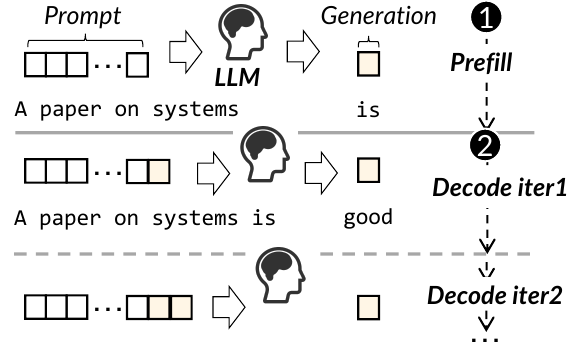} \\[1pt]    
        \end{minipage} \\[1pt]
        \begin{minipage}{1\linewidth}
        \caption{\small{\emph{
            An illustration showing how an LLM processes requests. 
        }}}
        \label{fig:bg-llm}
        \end{minipage} \\[-23pt]
        \end{figure}

\nospacestitle{LLM serving. \,}
Large language models (LLMs) process requests in an auto-regressive way,
as shown in {\fig{fig:bg-llm}}.
Given a request (or a batch of requests),
where each request contains multiple tokens (prompt),
the model first executes a prefill phase to generate the first token (\ding{182})
with a forward pass.
The model then enters a decoding phase (\ding{183})
that iteratively generating the remaining tokens with multiple forward passes.
The iteration ends after the module outputs an end-of-sequence (EOS) token.
During each iteration,
the input consists of both the original prompt and all previously generated tokens.

\begin{figure}[!t]
        \begin{minipage}{1\linewidth}
        \hspace{-1mm}
        \centering    
        \includegraphics[width=\columnwidth, trim=0.25cm 11.1cm 26.4cm 0.25cm, clip]{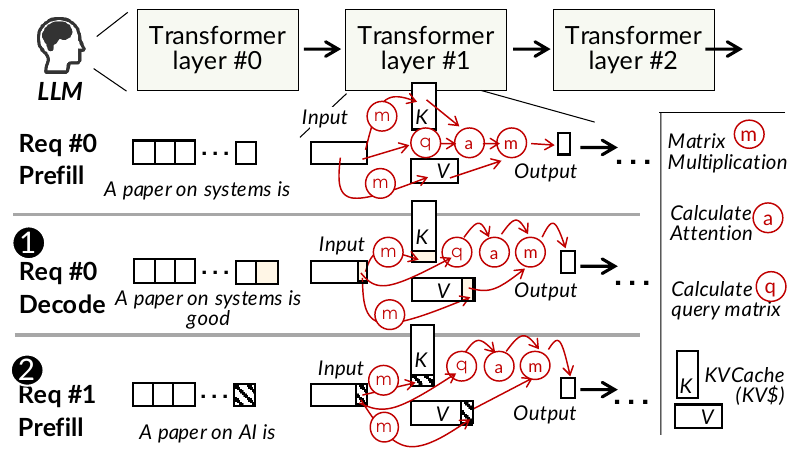} \\[1pt]    
        \end{minipage} \\[5pt]
        \begin{minipage}{1\linewidth}
        \caption{\small{ \emph{
        An illustration of: \ding{182} 
        how {\kvcache} from a prefill request (Req\#0) 
        can be reused by the decoding of Req\#0, 
        and \ding{183} how {\kvcache} can be reused for 
        the prefill of a future request (Req\#1). 
        }}}
        \label{fig:bg-kvcache}
        \end{minipage} \\[-10pt]
        \end{figure}

\stitle{KVCache (\kvcache). \,}
LLM performs three steps for a forward pass~\cite{DBLP:conf/nips/VaswaniSPUJGKP17} as shown in {\fig{fig:bg-kvcache}}:
(1) calculating the Q, K, and V matrices using matrix multiplication,
(2) using Q and K to derive attention scores,
and (3) combining the attention scores with the V matrix to produce the attention result.
These steps are computation-intensive.
Since inputs with the same tokens share the same K and V matrices (see \ding{182}),
all LLM serving systems cache the K and V matrices generated during the prefill phase
on the GPUs to save computation time.
These cached K and V matrices are then reused
during the decode phase for the same request to avoid recomputation, so they are termed as {\kvcache}.

\stitle{{\kvcache} cache and {\kvcache} blocks. \,}
Besides {\kvcache} reuse across prefill and decode phases
when processing a single request,
the {\kvcache} can also be reused across different requests
if they share the same token prefix.
As shown in {\fig{fig:bg-kvcache}} (\ding{183}),
the K and V values of ``A paper on''
are identical for both \#Req0 and \#Req1.
Thus, even when \#Req0 finishes decoding,
we can still cache its {\kvcache} for \#Req1,
which dramatically improves the reaction time (also termed time to first token, \emph{TTFT}) of \#Req1:
it only needs to compute the K and V of ``AI is'' during the prefill.
{\kvcache} caching further increases the serving throughput thanks to the saved computation.
Due to the above reasons, existing serving systems all cache the {\kvcache} of completed requests.
We term this approach {\kvcache} cache~\cite{vllm,cachedattention,promptcache,sglang,DBLP:journals/corr/abs-2404-12457,DBLP:conf/sigcomm/LiuLCRHZDY0AMHH24,DBLP:journals/corr/abs-2403-14401},
though other names like prefix cache or prompt cache also exist.
The {\kvcache} can be cached in GPU memory,
on the host memory (or remote machine's memory),
or even SSD~\cite{vllm,cachedattention,DBLP:journals/corr/abs-2403-14401}.

Since checking prefix-match may be costly at a per-token level,
the cache granularity of
existing {\kvcache} system is \emph{block},
where each block contains the KV of a configurable number of tokens.
For example, vLLM~\cite{vllm} by default groups 16 tokens as a block.

\stitle{Multi-turn serving. \,}
Interacting with LLMs through multi-turn conversations is a common pattern
to improve serving quality~\cite{wu2024autogen,wang2023mint}.
Specifically, after receiving the generation from the LLM,
the user may issue a next-turn request with a new prompt. 
Since the LLM needs the previous conversation context for processing, 
the serving system will concatenate the input and output of requests from the previous turns
to the new request's input, and feed the new input to the LLM for serving. 
To help manage the history of multi-turn requests, 
requests in a multi-turn conversation are typically grouped in a \emph{session}.
A multi-turn request naturally reuses {\kvcache} from previous turns 
since current LLM systems deliberately append the history of inputs and outputs 
as a prefix to each new request.

\section{Characterize {\kvcache} Workload in the Wild}
\label{sec:analyze}

\subsection{Trace data collection}
\label{sec:workload-data}

\begin{figure}[!t]
        \begin{minipage}{1\linewidth}
        \hspace{-1mm}        
        \centering    
        \includegraphics[width=\columnwidth, trim=0.25cm 11.9cm 26.6cm 0.25cm, clip]{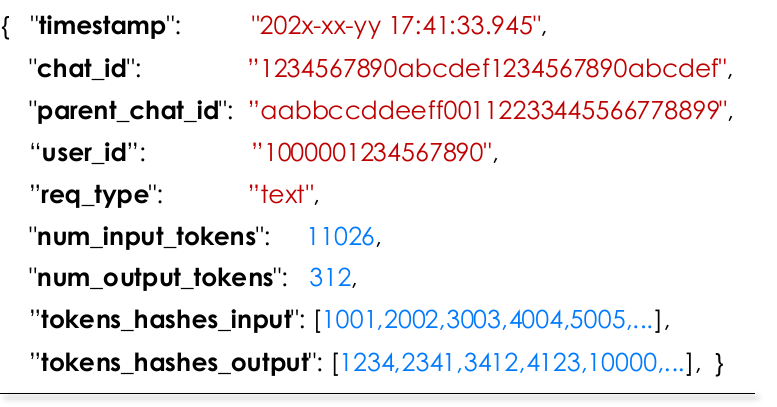} 
        \end{minipage} \\[5pt]
        \begin{minipage}{1\linewidth}
        \caption{\small{\emph{
            An example of the collected trace record. 
            }
        }}
        \label{fig:trace-record}
        \end{minipage} \\[-10pt]
        \end{figure}
        
\noindent
We collected and analyzed two weeks (February 2025 and December 2024)
of LLM serving request traces from
one of {\companyfull}'s serving clusters.
For brevity, we present results from one representative day per trace.
The results remain consistent throughout the workdays in each week (see \textsection{\ref{analysis:reuse-evolution}}).
\emph{Note that we cannot collect the raw input and output for each request
due to {\company}'s strict privacy policy.}
For a specific trace, all its collected requests are served by the same model.

Below, we first describe the detailed anonymized trace record provided by the operational team
of the serving cluster,
then we move on to highlight the differences between two traces,
both representing typical serving workloads nowadays.
{\fig{fig:trace-record}} presents the layout of a collected request:

\stitle{1. Timestamp: } The time 
when the serving cluster receives the request.  

\stitle{2. Chat ID: } The unique ID of the request. 

\stitle{3. Parent Chat ID (under multi-turn serving): } 
For requests in a multi-turn session,
the serving system assigns a parent ID to each request, linking it to the previous request in the session.
By traversing the parent IDs, we can identify all requests belonging to the same session. 
If the parent ID is empty, then the request is the first-turn of a session
and we call them \emph{single-turn request}. 
Otherwise, we call requests with a non-empty parent ID \emph{multi-turn request}. 

\stitle{4. User ID: }
The unique ID of the user submitting the request. 
These IDs are mapped to a randomly selected domain for anonymity. 

\stitle{5. Request type: } 
The serving gateway categorizes LLM requests into five types:
\emph{Text}: Plain text input via interfaces like ChatBot.
\emph{File}: file analysis (e.g., summarizing document content using LLM).
\emph{Multimodal}: Image understanding through OCR-enhanced processing.
\emph{Search}: Frontend with web-search integrated for knowledge augmentation.
\emph{API}: Programmatic access via OpenAI-compatible interfaces~\cite{openaiapi}.

\stitle{6. Number of Input/Output Tokens: }
The number of input tokens provided to the LLM
and the number of output tokens generated by it.

\stitle{7. The hashed values of In/Output tokens: }
Due to the strict privacy policy,
the runtime system generate the traces 
through salted hashing and domain remapping.
Specifically, the collector employs SipHash~\cite{SipHash} to
generate unique hashes for every four consecutive tokens with random salts per trace,
then remapping the hash IDs to consecutive natural numbers.
This fine-grained hashing enables more fine-grained cache analysis
while enhancing privacy protection compared to single-token hashing.

\stitle{Trace A vs. Trace B. \,}
We collected two traces (A and B) representing two representative LLM serving workloads.
Trace A represents a common to-C (customer) scenario where users interact
with the model through services built by the cloud,
i.e., ChatBot (Text) or File/Multimodal/Search services.
A key feature of to-C workloads is that they are typically human-in-the-loop.
Trace B represents a to-B (business) workload
where business users interact with the LLM by calling OpenAI-compatible APIs hosted by the cloud
through computer programs~\cite{openaiapi}.
API calling is critical for business users because they can systematically leverage
LLM to automate tasks like document translation or user request classification.
Trace B has no request type information because,
similar to the LLM interface calls currently offered by other vendors~\cite{openaiapi,genimiapi,claudeapi},
this information is concatenated on the client side.

\subsection{{\kvcache} reuses analysis in the wild}
\label{sec:workload-patterns}
            
\begin{figure}[!t]
    \centering
    \includegraphics[width=1.01\linewidth,center]{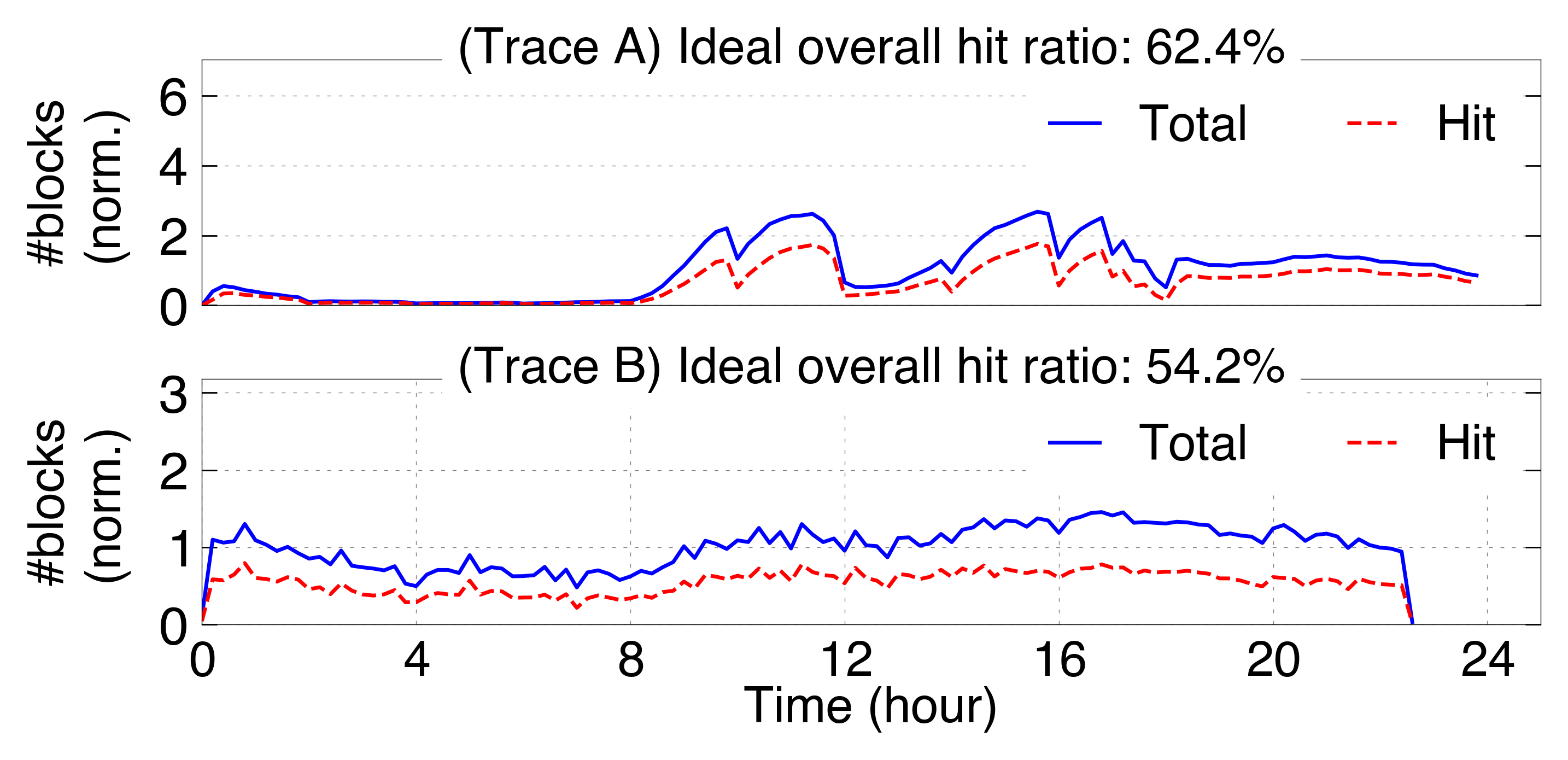} \\[2pt]
    \begin{minipage}{1\linewidth}
    \caption{\small{\emph{
        An analysis of 
        the ideal cache hit ratio of the {\kvcache} cache under real-world LLM serving workloads
        within a day. 
        The reported accessed and hit block numbers are normalized (norm.). 
    }}}
    \label{fig:motiv-cache-hit}
    \end{minipage} \\[-0pt]
\end{figure} 

\nospacestitle{How many {\kvcache} can be reused? \,} 
{\fig{fig:motiv-cache-hit}} shows the ideal cache hit rates
on two traces, i.e., assuming a cache having infinite capacity.
We can see that LLM workloads can benefit from caching---it has 62\% and 54\% hit rates
on Trace A and B, respectively.
The hit rate is calculated as follows:
for a total of $N$ {\kvcache} blocks that needed to be computed by LLM in a trace,
if the hit rate is 62\%, this implies that
62\% of the $N$ blocks can reuse a {\kvcache} computed and cached in a prior request.
While the ideal hit ratio is high,
it is smaller than the reported hit ratio
(e.g., more than 80\%~\cite{DBLP:journals/corr/abs-2403-14401,sglang,cachedattention})
on synthetic workloads.

\begin{figure}[!t]
        \begin{minipage}{1\linewidth}
        \hspace{-1mm}
        \centering    
        \includegraphics[width=.96\columnwidth, trim=0.25cm 12.54cm 26.9cm 0.25cm, clip]{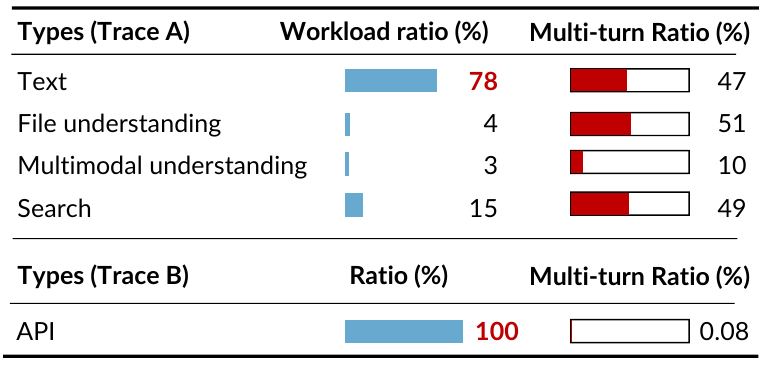} \\[1pt]
        \end{minipage} \\[5pt]
        \begin{minipage}{1\linewidth} 
        \caption{\small{\emph{%
            Workload types and multi-turn ratio of requests in our collected traces. 
        }}}
        \label{fig:motiv-workload-type}
        \end{minipage} \\[-10pt]
        \end{figure}  
  
\stitle{Single-turn and multi-turn requests are equally important for {\kvcache} reuse. \,}        
While it is a common belief that multi-turn requests are key contributors to the cache hits~\cite{DBLP:journals/corr/abs-2403-14401,cachedattention}, 
this is not the case on the to-B workloads (Trace B). 
{\fig{fig:motiv-workload-type}} shows 
the contribution of each request type as well as their multi-turn ratio.
For Trace A, it is dominated by the requests sent by the chatbot (78\,\%),
and all request types except for the multimodal have a similar multi-turn ratio (47--51\,\%).
For Trace B, it only has API requests
with a negligible multi-turn ratio (less than 0.1\,\%).
Nevertheless, the single-turn requests contribute 97\,\% of the total cache hits for Trace B.

Single-turn API requests dominate the hits in to-B workloads for two reasons. 
First, LLM API requests share {\kvcache} through common system prompts. 
Specifically, each request comprises:
(1) a system prompt providing task instructions (e.g., document translation guidelines) and 
(2) a user prompt containing request-specific data (e.g., target document). 
When programs make API calls~\cite{DBLP:conf/osdi/LinHZ00CQ24}, 
they typically hardcode identical system prompts in the program,
thus enabling {\kvcache} sharing across requests, even for single-turn requests.
Second, API requests often exhibit rapid submission rates (e.g., >\,10\,QPS),
enabling frequent reuse of the same shared {\kvcache} from system prompts. 

\begin{figure}[!t]
    \centering
    \includegraphics[width=1.01\linewidth,center]{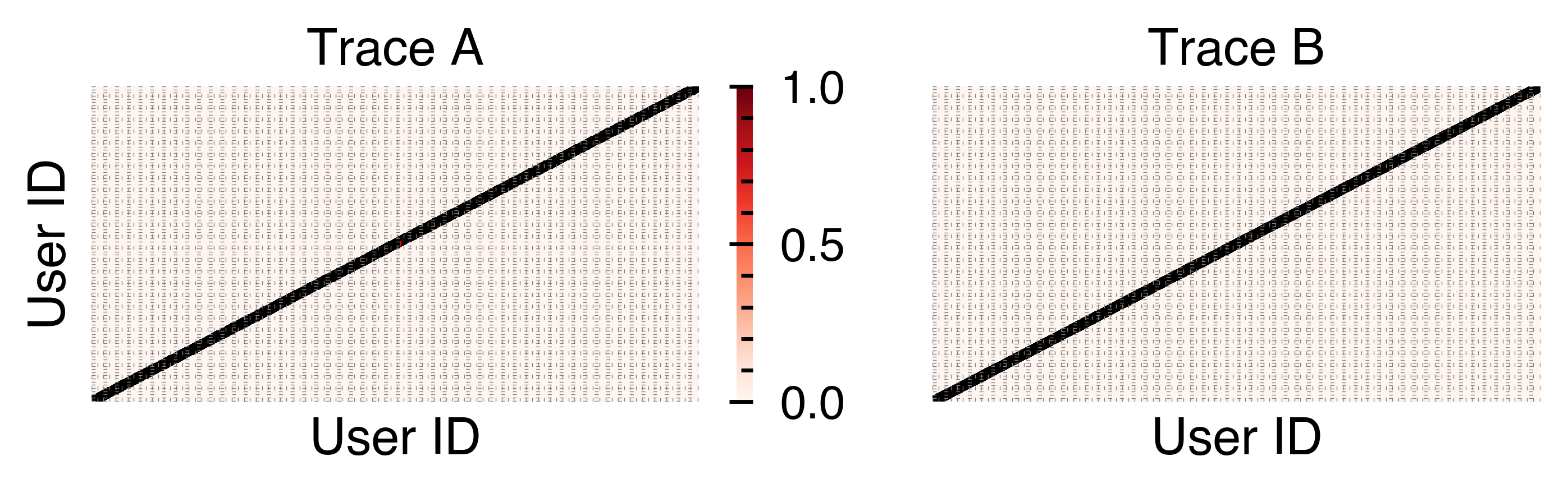} \\[-1pt]
    \begin{minipage}{1\linewidth}
    \caption{\small{\emph{
    An analysis of whether a user may hit the {\kvcache} cached by another user.
    The darker the color, the higher the normalized hit count.
    Note that we have amplified the diagonal line in case it is too thin, since
    the trace provider serves numerous users.
    }}}
    \label{fig:motiv-cross-hit}
    \end{minipage} \\[-5pt]
\end{figure} 

\stitle{Cross-user {\kvcache} hits are rare. \,}
{\fig{fig:motiv-cross-hit}} shows the heatmap of the cache hits between users,
assuming an infinite-size {\kvcache} cache.
Darker colors indicate higher normalized hit counts.
We can see that for each user, most hits come from requests submitted
by themselves (the diagonal line).
While users naturally reuse {\kvcache} from their own requests---either
via multi-turn conversations or through programs calling API with
the same system prompt for the same task, the low inter-user hit rate suggests that
users tend to customize their own system prompts rather than
using template prompts provided by third-party libraries~\cite{prompt-library}.
As a result, we suggest paying more attention to intra-user {\kvcache} 
reuses rather than inter-user~\cite{promptcache,sglang,DBLP:journals/corr/abs-2404-12457,DBLP:conf/sigcomm/LiuLCRHZDY0AMHH24}
when designing {\kvcache} systems.

\begin{figure}[!t]
    \centering
    \includegraphics[width=1.0\linewidth,center]{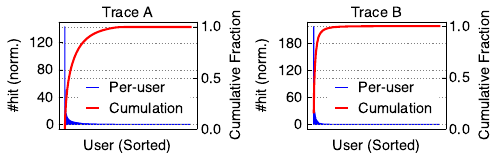}  \\[1pt]
    \begin{minipage}{1\linewidth}
    \caption{\small{\emph{        
    An analysis of the normalized (norm.) hits contributed by different users
    in an ideal setup with infinite cache size.
    }}}
    \label{fig:motiv-user-hit}
    \end{minipage} \\[-5pt]
\end{figure} 

\begin{figure}[!t]
    \centering
    \includegraphics[width=1.0\linewidth,center]{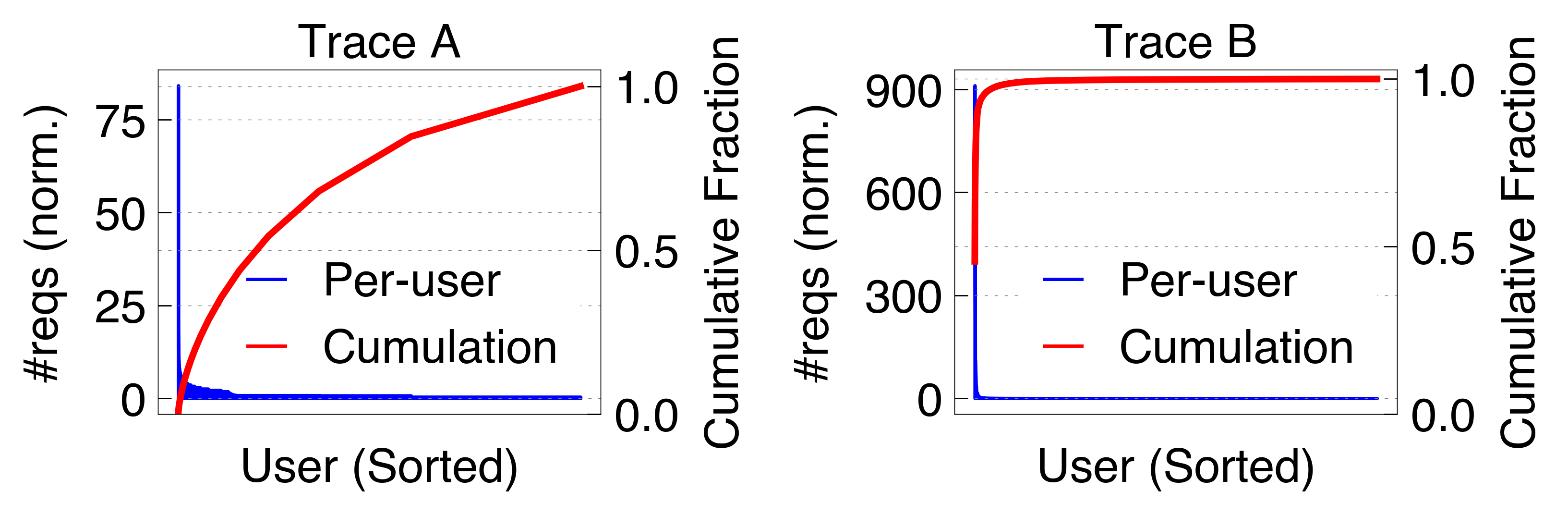}  \\[1pt]
    \begin{minipage}{1\linewidth}
    \caption{\small{\emph{        
        An analysis of the normalized (norm.) request count for different users. 
    }}}
    \label{fig:motiv-user-count}
    \end{minipage} \\[-5pt]
\end{figure} 

\stitle{The {\kvcache} reuses are skewed. \,}
{\fig{fig:motiv-user-hit}} shows the {\kvcache} cache hits contributed by different users,
along with the cumulative hits.
The {\kvcache} hit pattern is highly skewed: 19\,\% of the user requests in trace A and 4\,\%
in trace B account for more than 90\,\% of the {\kvcache} block hits.
This skewness stems from two factors.
First, the requests are unevenly distributed across users, especially in Trace B.
{\fig{fig:motiv-user-count}} shows the normalized number of requests sent by different users:
15\,\% of the users (head users) contribute 90\,\% of the requests in Trace B.
Since {\kvcache} hits primarily occur for requests from the same user (see {\fig{fig:motiv-cross-hit}}),
the skewness in request counts is expected to lead to a similar skewness in {\kvcache} hits.
Second, even when requests are sent relatively evenly by different users,
individual users vary significantly in their {\kvcache} hits.
As also shown in {\fig{fig:motiv-user-count}},
Trace A exhibits a moderately skewed request distribution: 19\,\% of users generate 50\,\% of the requests,
and 72\,\% generate 90\,\% of the requests.
Nevertheless,
the {\kvcache} hits are still skewed.
We suspect that this is due to their tendency to engage in multi-turn requests.

\subsection{Analyses of multi-turn requests features}
\label{sec:motiv-temporal-spatial}

\noindent
We first examine the features of multi-turn requests, as they 
still dominate the {\kvcache} reuse in to-C workloads. 
We omit the results on Trace B because multi-turn requests are rare in it.

\begin{figure}[!t]
    \centering
    \includegraphics[width=1.0\linewidth,center]{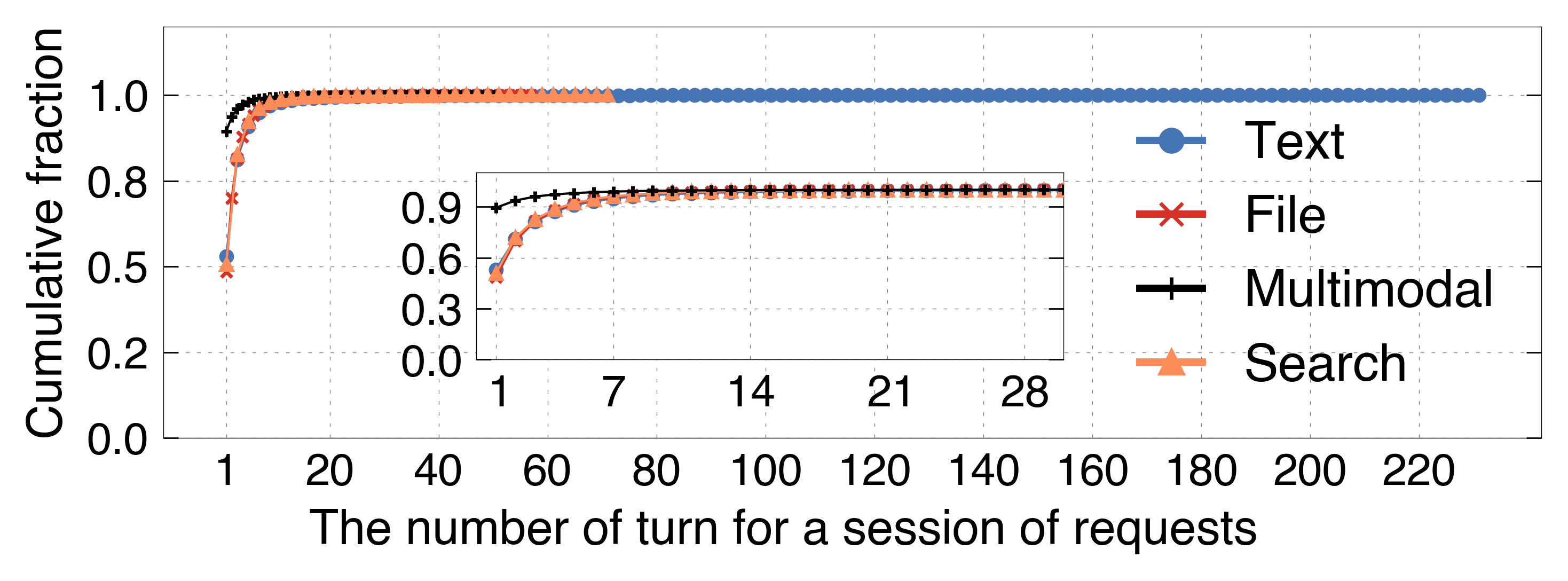}  \\[1pt]
    \begin{minipage}{1\linewidth}
    \caption{\small{\emph{        
        Distribution of turn number of multi-turn requests 
        on Trace A. 
    }}}
    \label{fig:motiv-multi-turn-numbersx}
    \end{minipage} \\[-5pt]
\end{figure} 

\stitle{Multi-turn requests have high variability. \,}
We observed two kinds of variability.
First, the number of turns varies significantly across different sessions. 
{\fig{fig:motiv-multi-turn-numbersx}} shows the distributions of turn numbers on Trace A: 
54\,\% of requests are single-turn requests, 
while the 50$^{th}$ and 90$^{th}$ percentile turn numbers are 1 and 5, respectively. 
The distribution has a long tail---the longest session has 232 turns, 
two orders of magnitude larger than the median (232\,$\times$).

\begin{figure}[!t]
    \centering
    \includegraphics[width=1.0\linewidth,center]{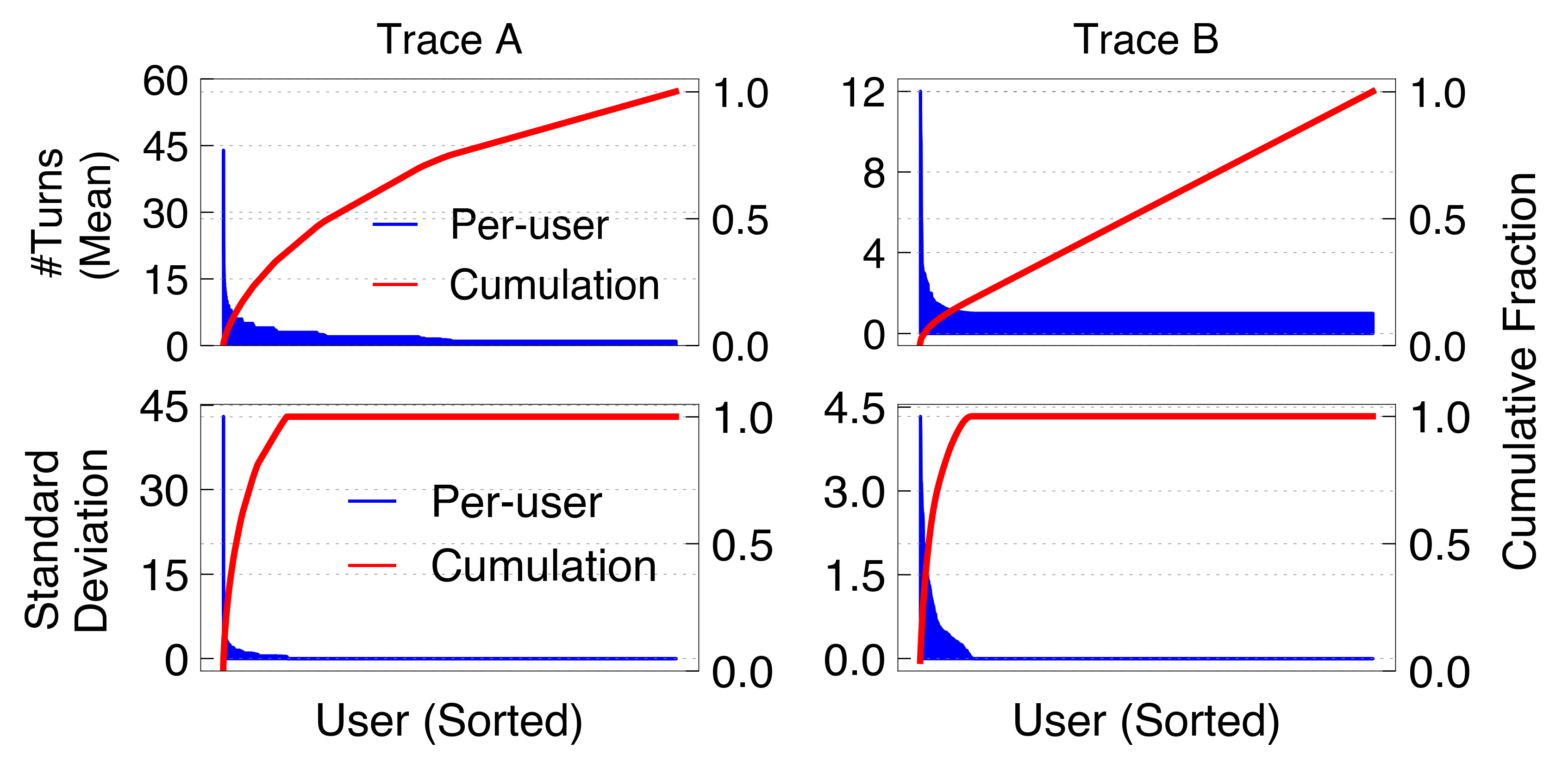}  \\[3pt]
    \begin{minipage}{1\linewidth}
    \caption{\small{\emph{        
        An analysis of the mean (upper) 
        and standard deviation (lower) of 
        the number of turns of requests sent by each user. 
        The users are sorted across to their mean number of turn (upper) 
        and standard deviation (lower).
    }}}
    \label{fig:motiv-user-turn-number}
    \end{minipage} \\[-5pt]
\end{figure} 

Second, multi-turn request distributions exhibit significant user-level variation.
{\fig{fig:motiv-user-turn-number}} shows that most users have similar mean turn counts,
except for a few head users—the top 10\% exhibit 8.9\,$\times$ more turns than the average.
Additionally, turn count deviations vary substantially, ranging from 1 to 44.
This suggests diverse interaction patterns: some users show high variability, while others remain stable.

\begin{figure}[!t]
        \begin{minipage}{1\linewidth}
        \centering    
        \includegraphics[width=.96\columnwidth, trim=0.25cm 7.9cm 24.1cm 0.25cm, clip]{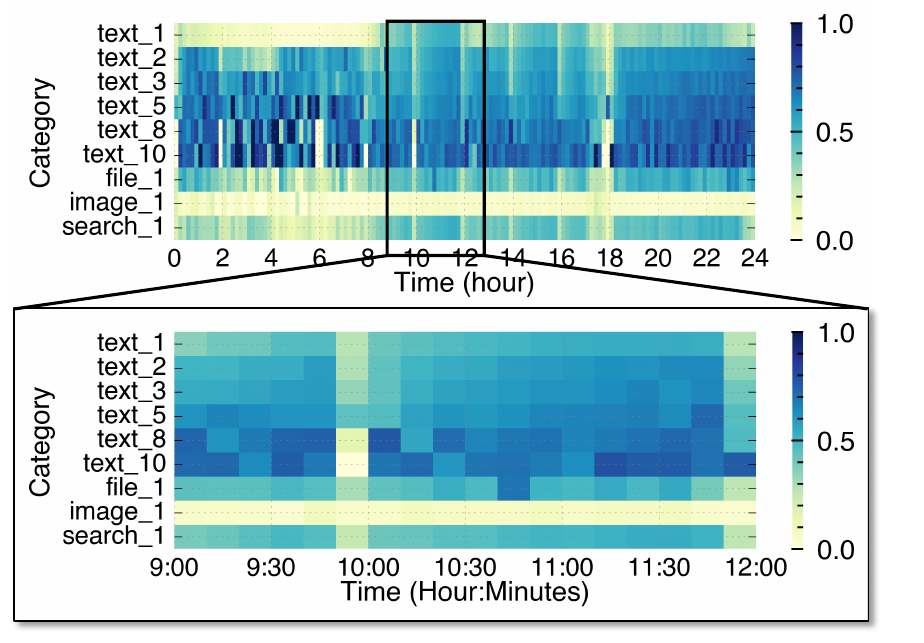} \\[1pt]
        \end{minipage} \\[5pt]
        \begin{minipage}{1\linewidth} 
        \caption{\small{\emph{%
            An illustration of how the frequencies of issuing the next turn request change over time 
        for different request types.
        The x-axis is the timeline.
        The workload categories are sorted according to their popularity.
        }}}
        \label{fig:motiv-next-turn-probability}
        \end{minipage} \\[-10pt]
        \end{figure}

\stitle{The probability of issuing the next turn request is predictable given a workload category and time interval,
and the probability differs across request categories. \,}
A request category includes both the original request type (e.g., Text and Search)
as well as the turn number of the request,
i.e., a 1-turn text (single-turn text request) is different from the 2-turn text request.
{\fig{fig:motiv-next-turn-probability}} shows a heatmap of how the
probability of issuing the next-turn request changes over the long time interval (24 hours)
as well as a zoomed-in view of a short time interval (3 hours).
The intensity of the color indicates
the profiled frequencies within a time window (10 minutes).
We can see that though different request categories have different frequencies (shown by vertical lines),
for a specific category,
the frequency of issuing the next-turn request remains consistent over a one-hour time interval
 (shown by horizontal lines).
We attribute this to the fact that the probability distribution of {\kvcache} reuse is quite fixed
given a specific time interval and request category.
\textsection{\ref{sec:motiv-temporal-spatial}} further elaborates on that.
Interestingly, the frequency of issuing the next-turn request differs across request categories.
For example, single-turn requests demonstrate significantly lower rates of 
issuing next-turn requests (mean=0.5) compared to multi-turn requests (mean=0.8).

\subsection{Analyses of temporal and spatial locality}
\label{sec:motiv-temporal-spatial}

\begin{figure}[!t]
    \centering
    \includegraphics[width=1.0\linewidth,center]{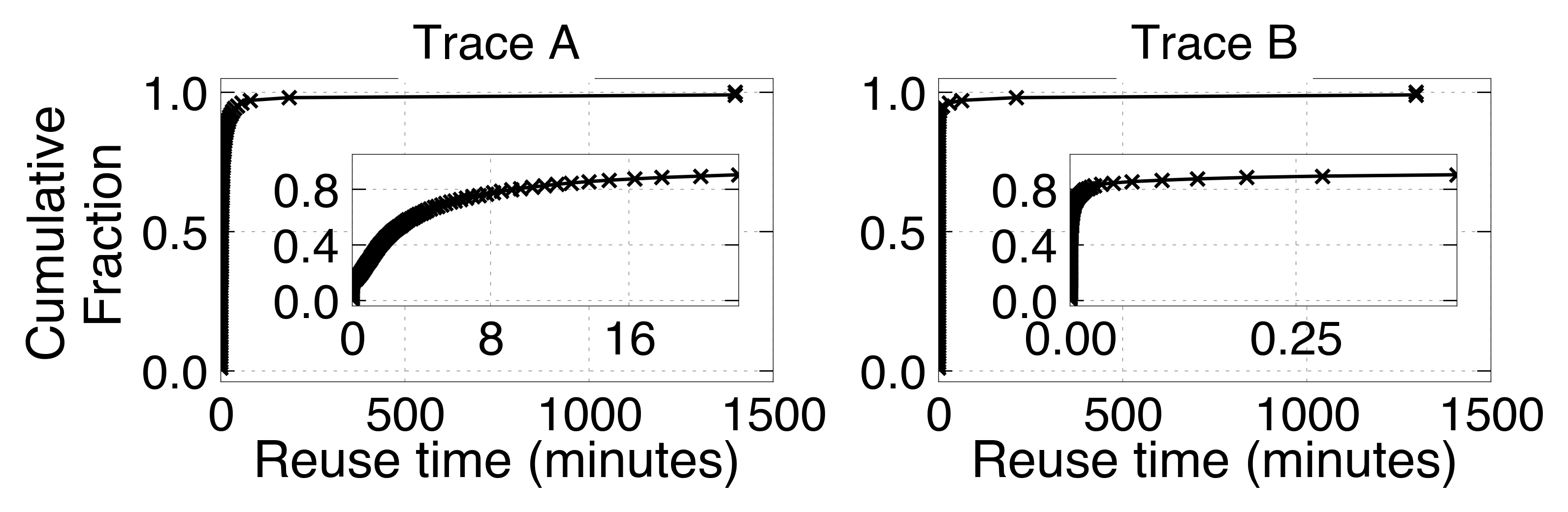}  \\[-0pt]
    \begin{minipage}{1\linewidth}
    \caption{\small{\emph{      
       The distribution of {\kvcache} reuse time of all requests in Trace A and B. 
    }}}
    \label{fig:motiv-temporal}
    \end{minipage} \\[-5pt]
\end{figure} 

\noindent
Temporal locality can be characterized by profiling the distribution of reuse time,
i.e., the interval between when a {\kvcache} block is reused by another request. 

\stitle{{\kvcache} reuse time is short. \,}
{\fig{fig:motiv-temporal}} shows the distributions of {\kvcache} reuse time
on both traces. 
We can see that the reuse time for both workloads is typically small:
In Trace A, 80\% of the reuse time falls within less than 10 minutes,
while in Trace B it falls within 10 seconds.
The differences lie in the human-in-the-loop versus computer-in-the-loop interactions
that generated the requests in two traces.

\begin{figure}[!t]
    \centering
    \includegraphics[width=1.0\linewidth,center]{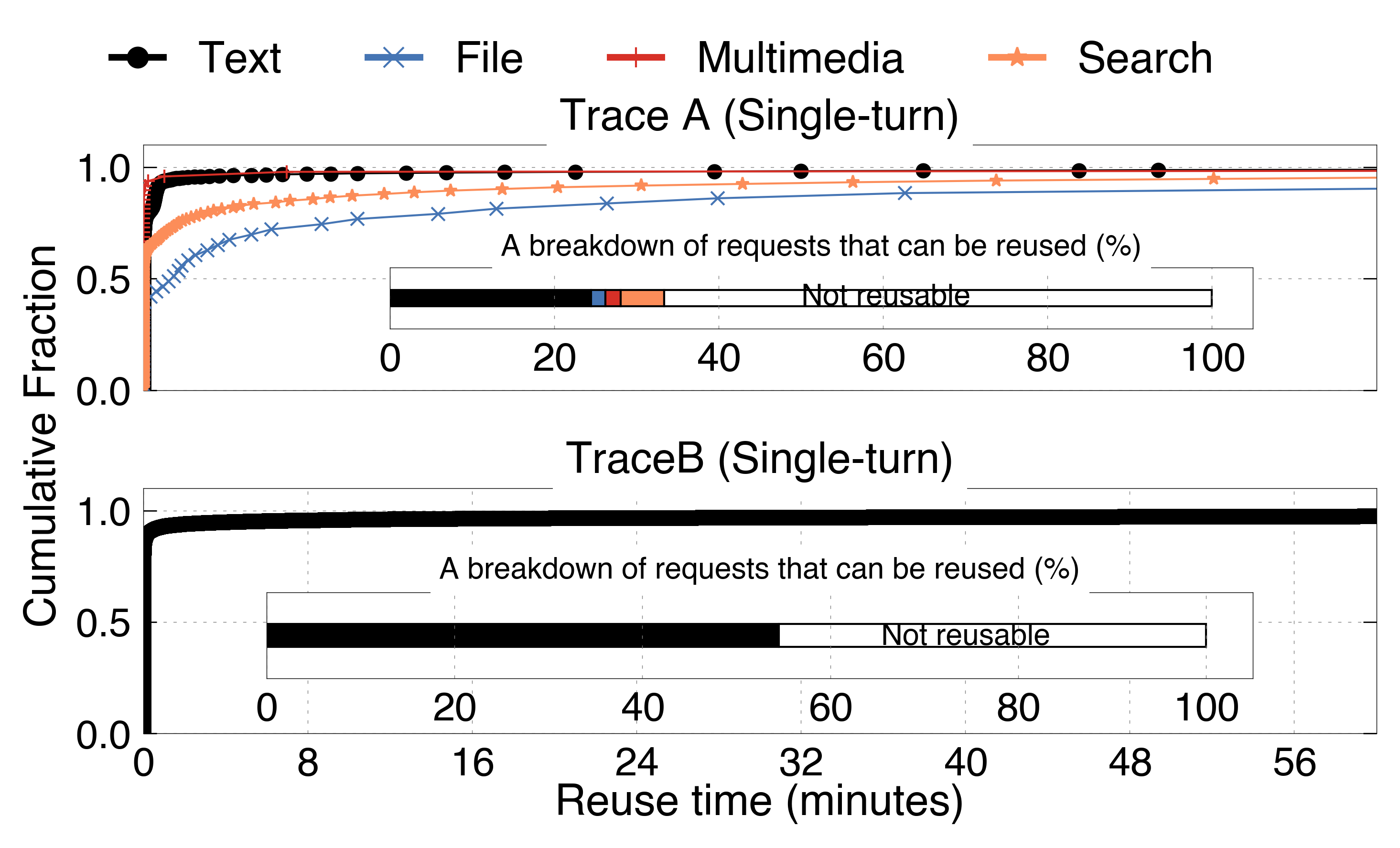}  \\[1pt]
    \begin{minipage}{1\linewidth}
    \caption{\small{\emph{        
        The distribution of {\kvcache} reuse time of single-turn 
        requests in Trace A and B. B has one category of request (API).
            }}}
    \label{fig:motiv-temporal-workload}
    \end{minipage} \\[-5pt]
\end{figure} 

\begin{figure}[!t]
    \centering
    \includegraphics[width=1.0\linewidth,center]{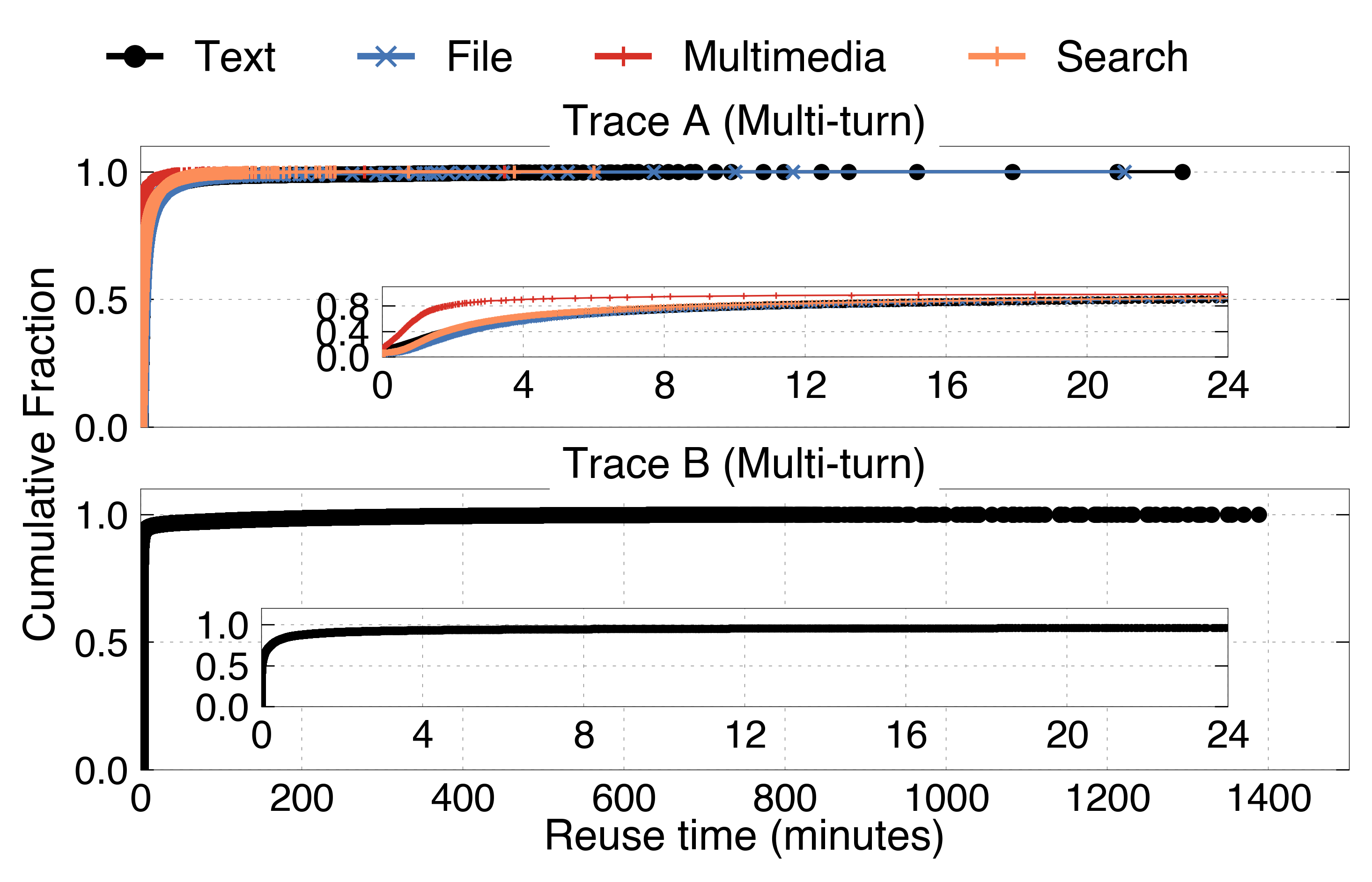}  \\[0pt]
    \begin{minipage}{1\linewidth}
    \caption{\small{\emph{        
        The distribution of {\kvcache} reuse time of multi-turn 
        requests in Trace A and B. B has one category of request (API).
        }}}
    \label{fig:motiv-time-between-turns}
    \end{minipage} \\[-5pt]
\end{figure} 

\stitle{Different workload types have different temporal locality. \,}
First, single-turn requests have much lower locality in the to-C workloads, 
while to-B workloads are the opposite. 
{\fig{fig:motiv-temporal-workload}} reports 
whether a single request can be reused by another single-turn request on both traces
in the middle sub-figures.
As we can see, less than 30\,\% of the single-turn requests are reusable in Trace A, 
but more than 50\,\% of requests can be reused in Trace B. 

Second, different request types show different temporal locality.
{\fig{fig:motiv-time-between-turns}} shows the reuse time distribution of multi-turn requests
categorized by their request types.
We can see that chat requests (Text) shows longer reuse time than 
image understanding (Multimodal), but shorter than file understanding (File).
This stems from workload characteristics:
users typically spend more time processing complex LLM outputs based on file content
than casual outputs from conversational chats.
Conversely, visual information processing occurs faster than textual processing~\cite{rock2010your},
leading to shorter reuse times for image tasks.
Finally, Search workloads exhibit the second-longest reuse time,
as users often require substantial time to analyze the searched results.

\begin{figure*}[!t]
    \centering
    \includegraphics[width=1.01\linewidth,center]{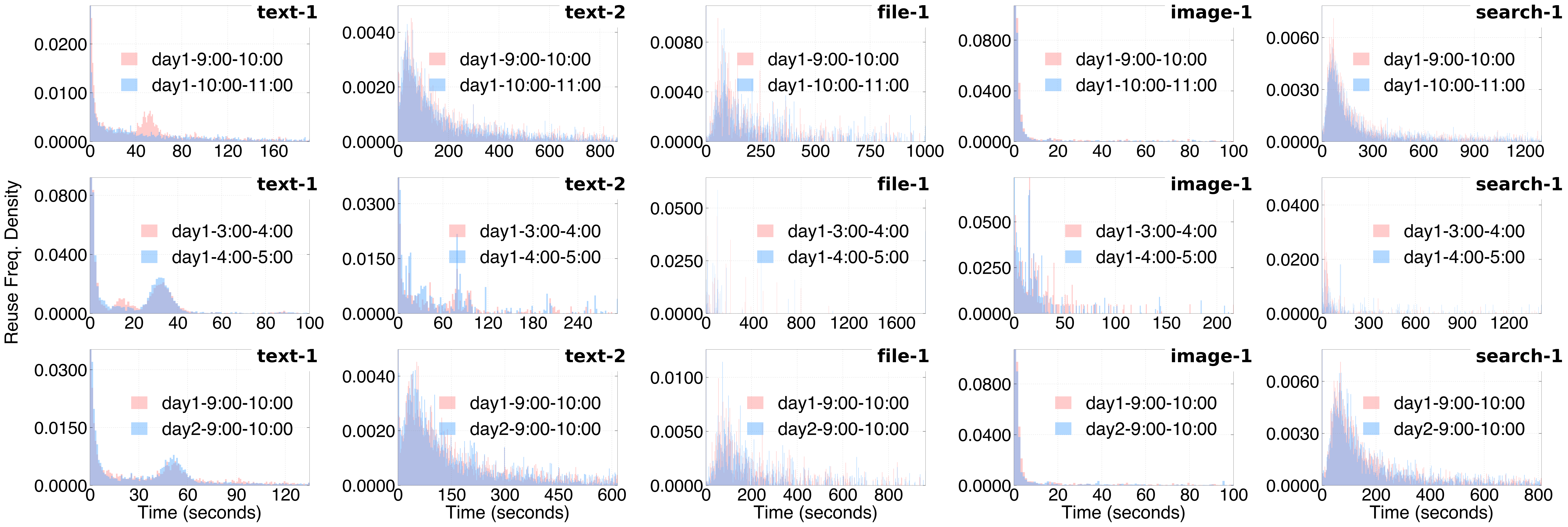}  \\ [6pt]
    \begin{minipage}{1\linewidth}
    \caption{\small{\emph{       
    An empirical analysis of the reuse time probability distribution of {\kvcache} blocks.
    The columns present data from different request categories.
    The first row: distributions under heavy traffic (daytime).
    The second row: distributions under low traffic (nighttime).
    The third row: a comparison of distributions under different days but with similar traffic patterns.
    }}}
    \label{fig:motiv-reuse-time-probability-hist}
    \end{minipage} \\[-5pt]
\end{figure*} 

\stitle{Given a time period and request category, 
the probability distribution of {\kvcache} reuse time follows an exponential distribution. \,}
\label{analysis:block-reuse-probability}
{\fig{fig:motiv-reuse-time-probability-hist}} presents the reuse time probability distribution of {\kvcache} blocks across different request categories. 
For each category, we measure the likelihood of its {\kvcache} reuse at specific time intervals after caching. 
We plot these distributions by counting reuse event frequencies at each second following request execution.

From the results, we draw three key observations. 
First, exponential distributions fit well with the reuse probability given
a request category. 
Second, the probability is workload-aware,
i.e., the distributions of different request categories,
including the same request type but different turns,
are different (e.g., text-1 vs. text-2). This implies a workload-aware design that distinguishes different request categories.
Finally, the distributions of requests with similar traffic patterns
are similar. Take text-1 as an example: as shown in the first row,
the distribution drawn from 9:00--10:00 a.m. is similar to that drawn from 10:00--11:00 a.m.,
but is significantly different from that drawn from the nighttime---3:00--4:00 a.m. (second row).
The distributions are also similar across different days (first row vs. third row).
The first and third observation in combine imply that we can acquire the distribution of a request category
using recent historical data.

\begin{figure}[!t]
    \centering
    \includegraphics[width=1.0\linewidth,center]{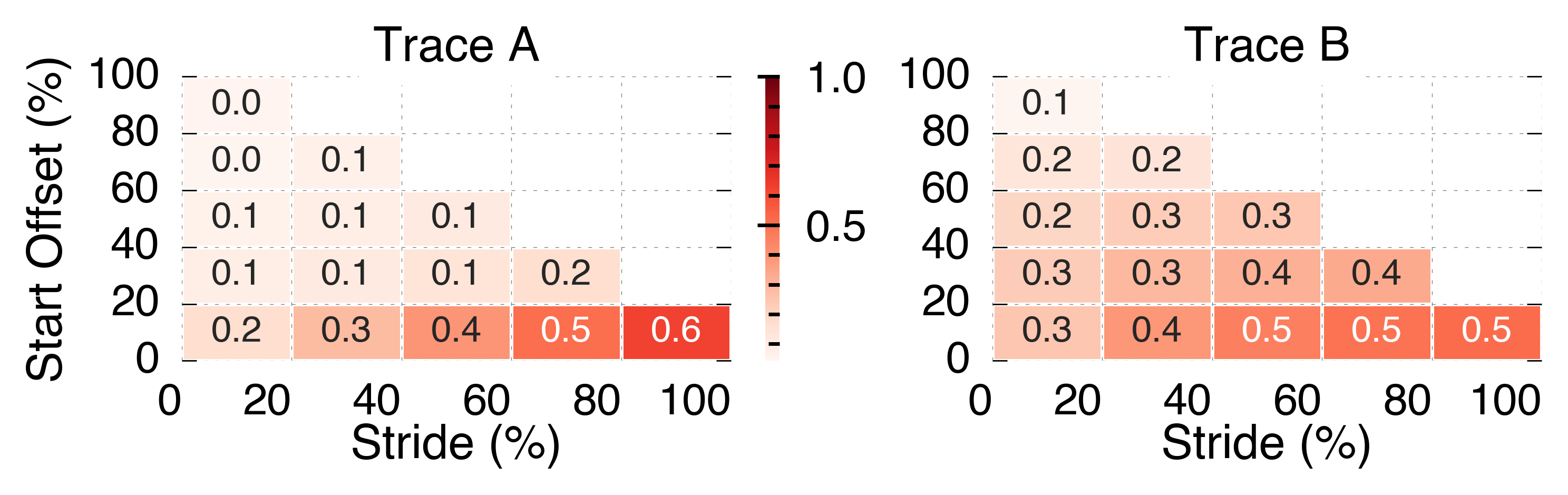}  \\[-3pt]
    \begin{minipage}{1\linewidth}
    \caption{\small{\emph{        
      Overall spatial locality of two traces.
    }}}
    \label{fig:motiv-spatial}
    \end{minipage} \\[-5pt]
\end{figure} 

\begin{figure}[!t]
    \centering
    \includegraphics[width=1.0\linewidth,center]{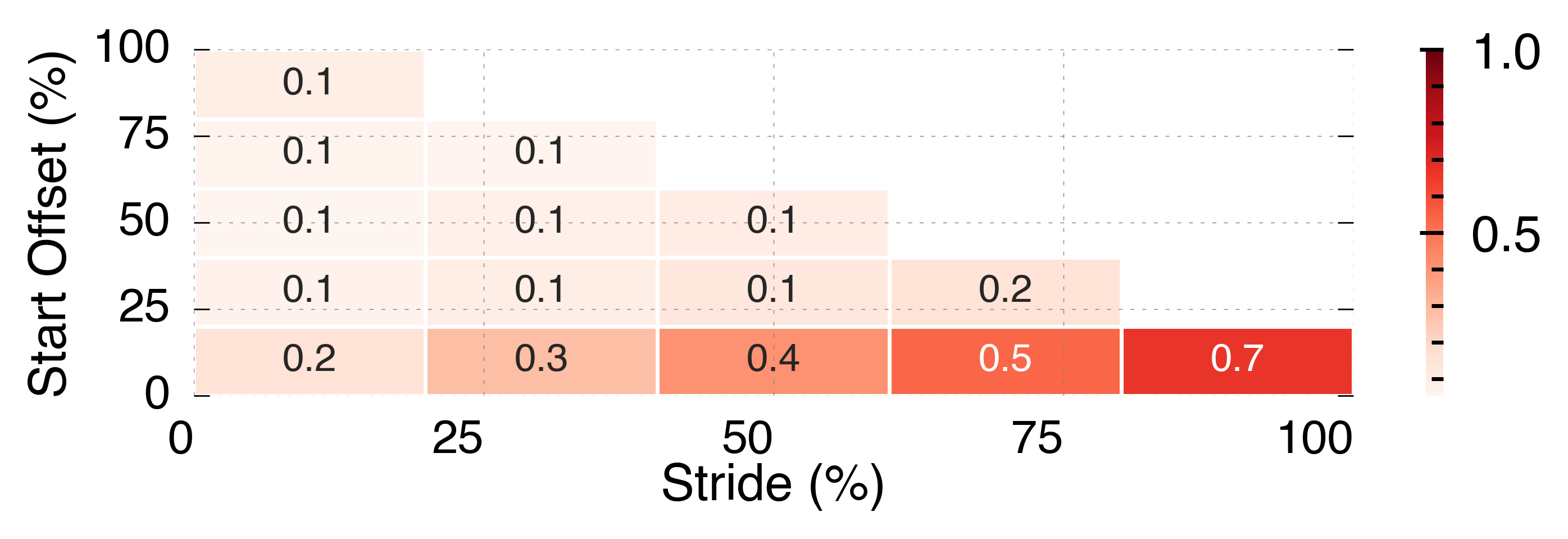}  \\[-5pt]
    \begin{minipage}{1\linewidth}
    \caption{\small{\emph{        
        The spatial locality of multi-turn requests on Trace A.
    }}}
    \label{fig:motiv-spatial-multi}
    \end{minipage} \\[-5pt]
\end{figure}

\stitle{Spatial locality. }
{\fig{fig:motiv-spatial}} shows the overall spatial locality of {\kvcache} across both traces.
We characterize the spatial locality by measuring the ideal cache hit ratio
with various cache strides, i.e., the percentage of tokens cached per request,
as well as different starting token offsets for caching.
This gives a two-dimensional heatmap,
for example, offset (10\,\%, 10\,\%) means that for each request,
we only cache 10\,\% of the tokens starting from the 10\,\% token position.

First, {\kvcache} shows better spatial locality when caching from the beginning,
which is expected since only requests with identical prefixes can share the {\kvcache} (\textsection{\ref{sec:bg}}).
We emphasize this seemingly trivial point because some {\kvcache} optimizations~\cite{DBLP:journals/corr/abs-2410-03065} cache
the tail tokens of requests to hide {\kvcache} load time through concurrent computation and loading.
However, this approach may reduce spatial locality,
as {\fig{fig:motiv-spatial-multi}} particularly demonstrates that reusing intermediate parts
in the cache yields minimal benefits for multi-turn conversations.

\begin{figure*}[!t]
    \centering
    \includegraphics[width=0.95\linewidth,center]{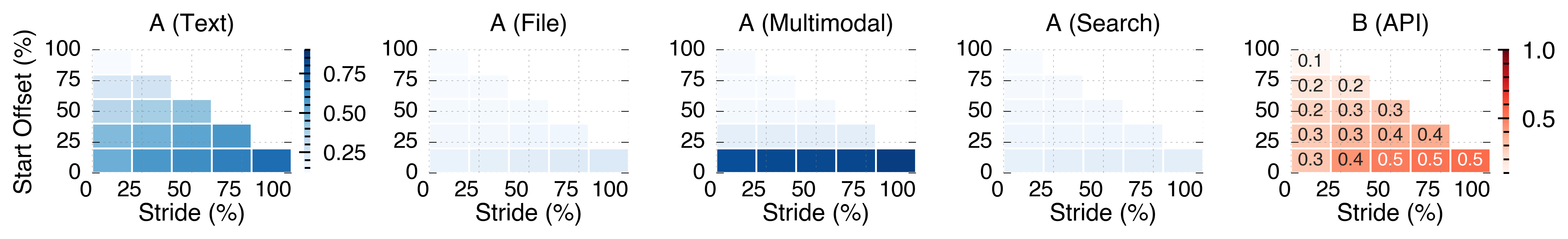} \\[-4pt]
    \begin{minipage}{1\linewidth}
    \caption{\small{\emph{
        The spatial locality of single-turn requests. 
    }}}
    \label{fig:motiv-spatial-workload}
    \end{minipage} \\[-5pt]
\end{figure*} 

Second, the spatial locality, like temporal locality, is also workload-aware.
In {\fig{fig:motiv-spatial-workload}}, we observed workloads
like text and multimodal exhibit obvious spatial locality,
while file and search exhibit almost no spatial locality.
Due to the lack of user input content,
we suspect the cause as follows:
the applications add a fixed system prompt for text and multimodal workloads,
whereas the system prompt for file and search is dependent on user inputs.

Finally, we found that increasing the stride size implies a higher cache hit ratio in Trace A ({\fig{fig:motiv-spatial-multi}}),
but it is not the case in Trace B (the last column in {\fig{fig:motiv-spatial-workload}}).
In Trace A, multi-turn requests contribute to the cache hits,
so caching more tokens for a request will lead to more cache hits.
In Trace B, only the system prompt contributes to the cache hits.
Thus, caching a stride of 50\,\% of the request already achieved a peak hit ratio of 50\,\%.
This implies that most system prompt lengths are below 50\% of the total request length in Trace B,
though we don't have the detailed request to support this suspicion.

\subsection{Analyses of {\kvcache} cache capacity requirement}
\label{sec:analyze-capacity}

\noindent
Determining the right cache capacity is critical for a caching system.
Unlike traditional storage systems,
the cached {\kvcache} is ephemeral, so identifying the right size requirement
can save platform costs.
This section analyzes the {\kvcache} capacity using a bottom-up approach:
First, we characterize the features of the lifespan of requests' {\kvcache}.
Then we analyze the storage usage of each request.
In combination, this gives us an estimation of the total {\kvcache} capacity required.


\begin{figure}[!t]
    \centering
    \includegraphics[width=1.0\linewidth,center]{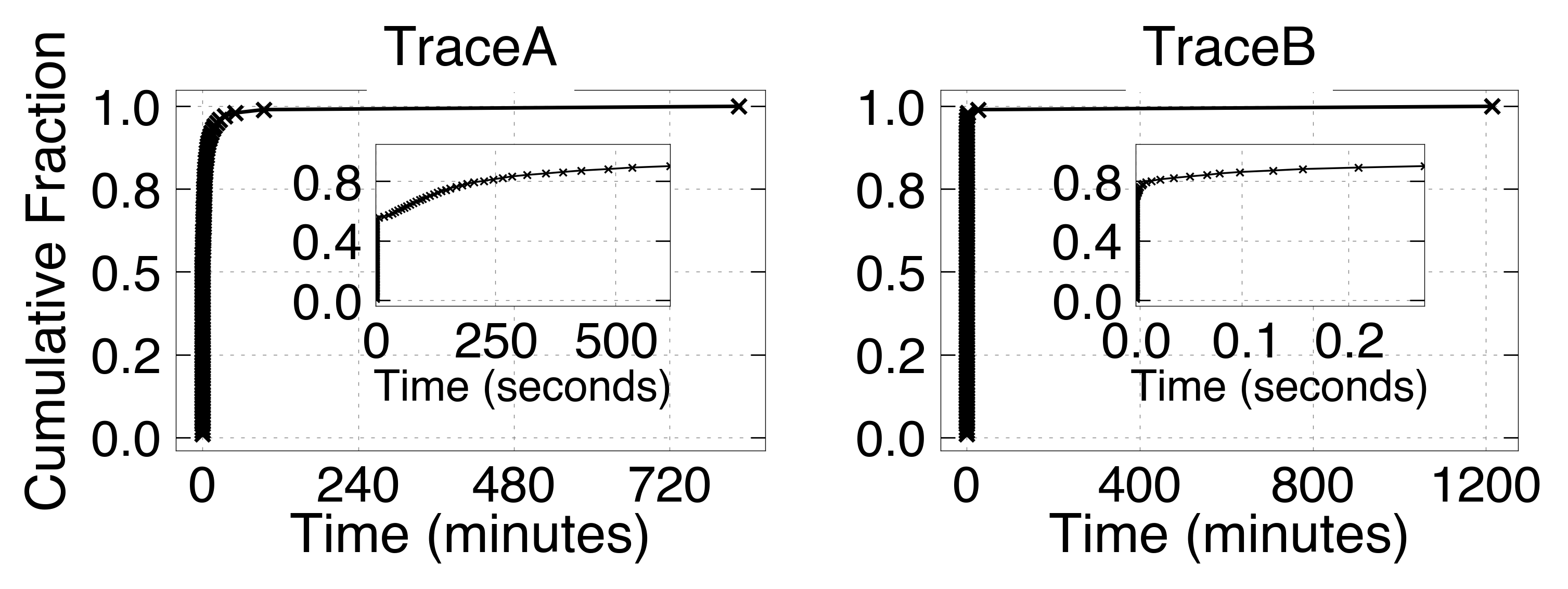}  \\[0pt]
    \begin{minipage}{1\linewidth}
    \caption{\small{\emph{        
        The distribution of the lifespan of {\kvcache} blocks on both traces. 
    }}}
    \label{fig:motiv-block-timespan}
    \end{minipage} \\[-5pt]
\end{figure} 

\stitle{{\kvcache} lifespan is ephemeral and predictable. }
{\fig{fig:motiv-block-timespan}} analyzes the lifespan of {\kvcache} blocks.
A {\kvcache} block ``dies'' if it has not been reused anymore.
We can see that most {\kvcache} blocks ``live'' shortly.
In Trace A, 90\,\% of the {\kvcache} blocks are not reused after 612 seconds,
and the number for Trace B is 0.3 seconds.
This aligns with our analysis in the previous sections,
e.g., many multi-turn requests will have a short reuse time ({\fig{fig:motiv-temporal}}),
and most requests have a few turns ({\fig{fig:motiv-multi-turn-numbersx}}).
As a multi-turn request ends, its {\kvcache} will not be reused anymore.
Blocks in Trace B have a shorter lifespan despite some outliers.
Unlike Trace A,
most of its cached {\kvcache} blocks are reused for system prompts,
so they can be reused after a long time.
For example, in Trace B, the longest lifespan is 1,200 minutes
while it is 800 minutes in Trace A.

\begin{figure}[!t]
    \centering
    \includegraphics[width=.96\columnwidth, trim=0.25cm 7.85cm 24.1cm 0.25cm, clip]{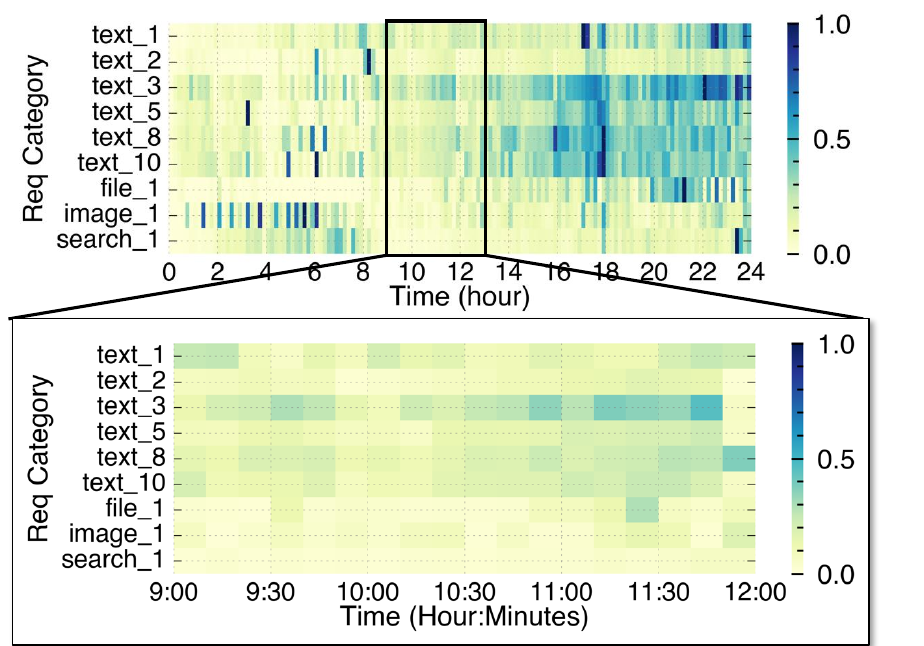} \\[5pt]
    \begin{minipage}{1\linewidth}
    \caption{\small{\emph{        
        An illustration of how the mean lifespan of 
        {\kvcache} blocks changes over time for different request categories. 
        Note that the x-axis represents the timeline.
        The request categories are sorted according to their usage.
    }}}
    \label{fig:motiv-block-timespan-probability}
    \end{minipage} \\[-5pt]
\end{figure}


{\fig{fig:motiv-block-timespan-probability}} further shows how the mean lifespan of {\kvcache} blocks
changes over time for different request categories.
In the graph, the color intensity represents the normalized lifespan,
with darker colors indicating longer-lived {\kvcache} blocks.
We can see that although the average lifespan at the 24-hour scale is changing over time,
when we zoom in to a smaller timescale (9:00 a.m. to 11:00 a.m.),
the pattern is quite stable.
This implies that we can predict the lifespan of a {\kvcache} block
using historical data to reduce the {\kvcache} cache size
for saving costs,
as storing {\kvcache} at fast storage medium (e.g., CPU's DRAM) is expensive. 

\begin{figure*}[!t]
    \centering
    \includegraphics[width=1.01\linewidth,center]{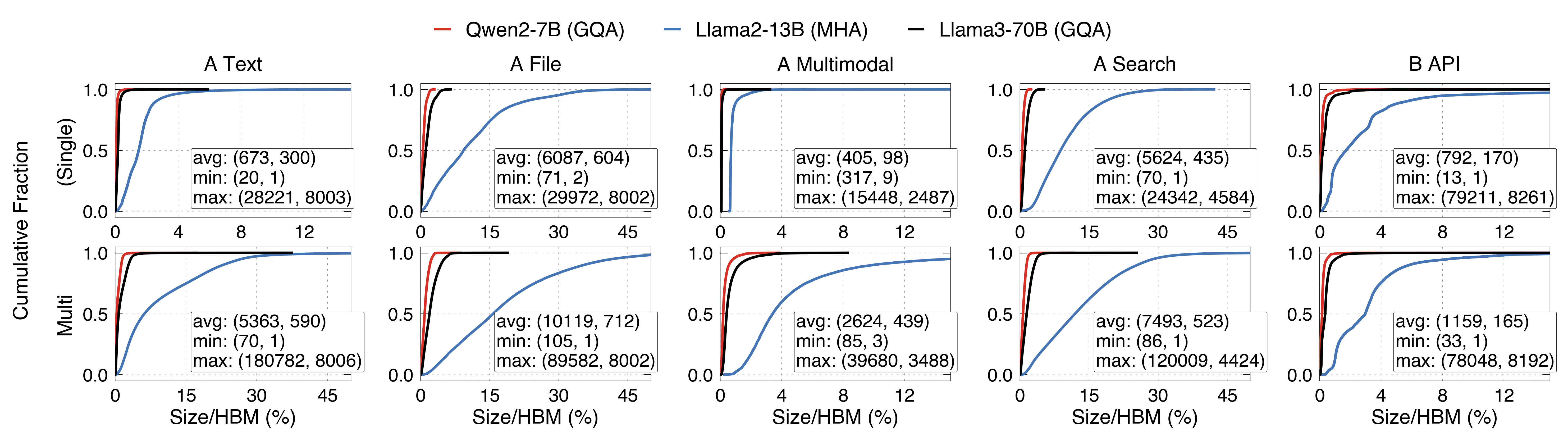} \\[2pt]
    \begin{minipage}{1\linewidth}
    \caption{\small{\emph{
        The distribution of {\kvcache} size of requests on different traces for \uline{S}ingle 
        and \uline{M}ulti-turn requests. 
        The size is normalized to the 
        GPU memory (HBM) available for serving. 
        The number in the tuple inside each graph represents the length of input and output.
    }}}
    \label{fig:motiv-cache-size}
    \end{minipage} \\[-0pt]
\end{figure*}

\stitle{Per-request {\kvcache} cache usage is moderate. }
A key factor when considering caching is the size of {\kvcache} to be cached,
which is proportional to the input length of each request and
the corresponding model output length.
{\fig{fig:motiv-cache-size}} shows the size distribution of requests on different traces.
Note that we report the relative size when normalized to the HBM available for serving the request
on a serving instance\footnote{\footnotesize{A serving instance
is the minimal number of GPUs required to serve a model.
For example, the serving instance of an 8\,B model only needs one GPU,
while serving a 72\,B model requires 4 in the common case. }}.
This is because a relative value
gives more intuitive results on the capacity required to cache
the {\kvcache} of each request.
The available HBM for {\kvcache} is calculated by subtracting the total HBM of the GPU(s)
in a serving instance from the model parameter and activation sizes.
Since the {\kvcache} size of each request depends on the detailed attention mechanism of the model,
we report the results on three representative models with two widely used attention methods:
group query attention (GQA~\cite{DBLP:conf/emnlp/AinslieLJZLS23}) and multi-head attention (MHA~\cite{DBLP:conf/nips/VaswaniSPUJGKP17}).
GQA uses less memory for the {\kvcache}.

We can see that the overall size of the {\kvcache} is moderate compared to the HBM.
Take Trace A as an example.
On GQA models like Qwen2-7B,
the 50$^{th}$, 90$^{th}$, and 99$^{th}$ HBM usage of the request in Text---the dominant request type
are 0.09\,\%, 0.15\,\%, and 0.37\,\% and 0.36\,\%, 1.31\,\%, and 2.21\,\% for single-turn and multi-turn requests,
respectively.
The average usage is 0.09\,\% and 0.56\,\% for single- and multi-turn requests, respectively.
This small usage is due to the fact that each token consumes few megabytes and
requests in production traces tend to be small:
in Qwen2-7B, it consumes 0.875\,MB {\kvcache} for 16 tokens,
while the mean total token lengths (input+output) for single-turn and multi-turn is 973 and 5,953, respectively.
For other models, though Llama3-70B also adopts GQA,
it has a slightly higher HBM usage due to the reduced number of GPUs per instance: the
parameter size increases 10\,$\times$ while the number of GPUs we measured only increases 4\,$\times$.
Finally, MHA models have the highest usage due to the increased {\kvcache} per token.
Note that recent open-source models tend to use GQA (or other
attention with even fewer {\kvcache} per token like MLA~\cite{liu2024deepseek}).

Considering that each serving instance typically processes a few requests per second,
our characterization implies that an instance can cache a request for a modest time---e.g.,
for Qwen2-7B, suppose a GPU can process 5 requests per second,
it takes at least 1 minute to saturate the HBM assuming an average 0.32\,\% HBM per request (derived
with a 47\,\% multi-turn ratio of Text request type).
This time is within the reuse distance of {\kvcache} in many scenarios (e.g., API).
Note that in practical setups, we may not achieve such a long reuse time
due to tail requests (e.g., we do observe long context requests with more than 12,000 tokens),
and LLM serving systems typically serve requests in large batches to improve computation utilization.

\begin{figure}[!t]
    \centering
    \includegraphics[width=1.0\linewidth,center]{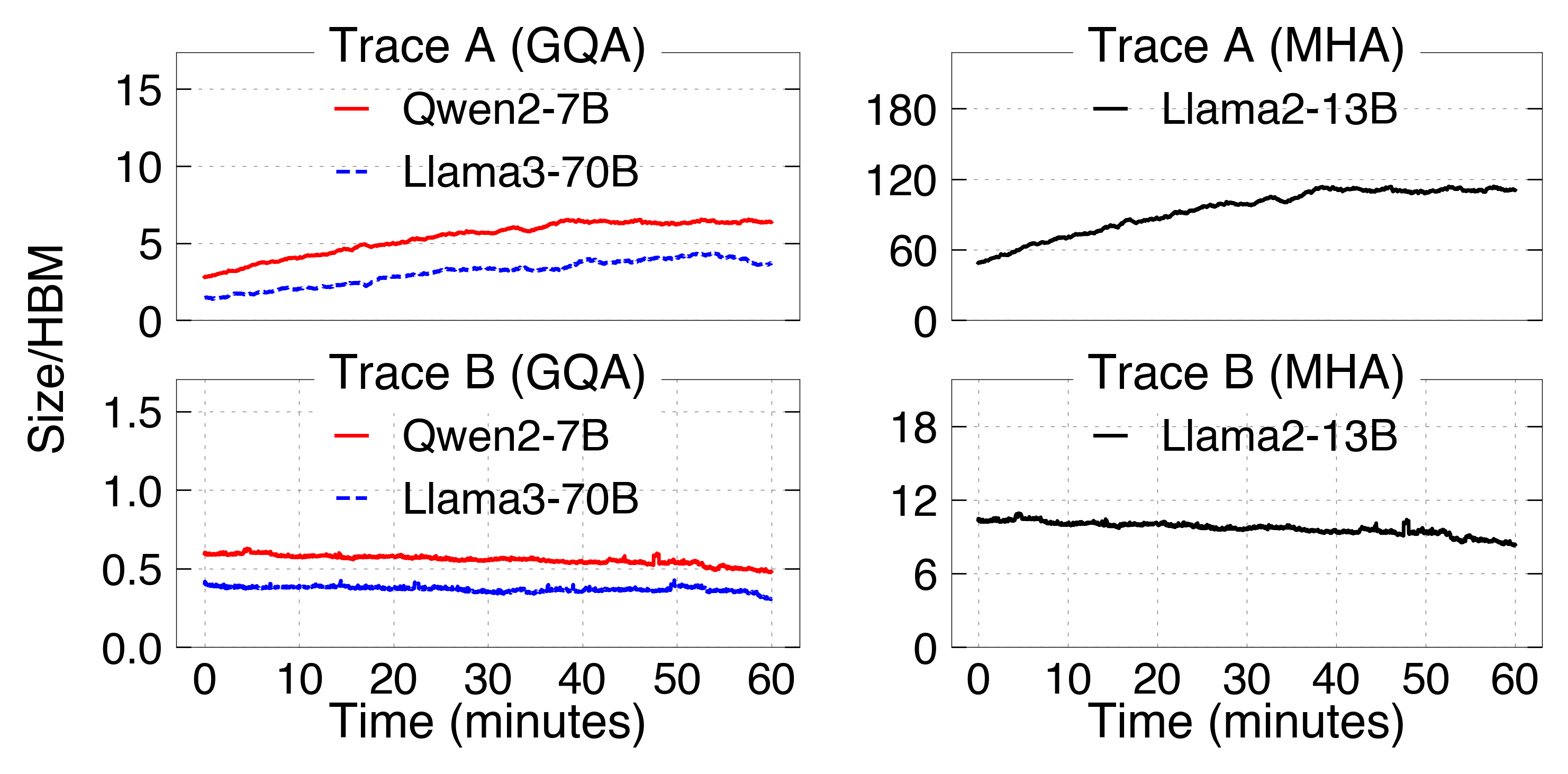}  \\[3pt]
    \begin{minipage}{1\linewidth}
    \caption{\small{\emph{        
    An analysis of the {\kvcache} cache size required
    to achieve an ideal cache hit ratio on different models and traces.
    The data is collected from 10:00 a.m. to 11:00 a.m.
    }}}
    \label{fig:motiv-opt-cache-space}
    \end{minipage} \\[-5pt]
\end{figure} 

\stitle{Overall memory required for {\kvcache} cache is moderate. }
We analyze the {\kvcache} cache capacity required
by measuring how large a cache can ensure an ideal cache hit ratio
assuming an ideal eviction policy,
i.e., never evict a {\kvcache} that will be reused.
The measurement is as follows:
at a given time, upon processing a request,
we update the {\kvcache} size by adding the {\kvcache}
of this request if it will be reused,
and subtracting the size of all cached {\kvcache} that will never be reused.
One tricky thing is that
the size required is dependent on the number of serving instances deployed for a workload:
if the instances are over-provisioned, then the estimated requirement would be small.
To obtain the maximal requirement,
we use a setup that uses the minimal number of serving instances,
i.e., each instance serves with the maximum QPS,
so there is no over-provisioning.

{\fig{fig:motiv-opt-cache-space}} shows the results normalized to the
total HBM of a serving instance.
We can see that GQA models require a relatively small cache capacity to reach an ideal cache hit rate:
in Trace A, Llama3-70B only needs 4\,$\times$ of the available HBM for {\kvcache}.
This is typically achievable in modern serving clusters without the need for more storage hierarchies
like remote memory or SSD~\cite{cachedattention}.
For example, in a typical setup featuring 8 A100 GPUs and 1\,TB of CPU memory,
each GPU has access to an average of 128\,GB of CPU memory,
which is nearly 4\,$\times$ the amount of HBM space reserved for storing the {\kvcache}.
Moreover, for Trace B, the {\kvcache} capacity is even smaller than the reserved HBM,
so caching on GPU may be sufficient.

Though GQA needs small cache, 
we should note that MHA models do require a huge amount of {\kvcache} cache,
This is due to their huge per-token KV pairs (see {\fig{fig:motiv-cache-size}}). 
Thus, optimizing cache eviction policies is still important under 
limited cache capacity.

\section{Improved {\kvcache} cache system}
\label{sec:design}

\subsection{An overview of the design space}
\label{sec:design-mechanism}

\noindent
Existing {\kvcache} cache systems~\cite{cachedattention, vllm,sglang,DBLP:journals/corr/abs-2403-14401}
follow similar caching mechanisms and policies.
Below we give a brief overview of their designs
before moving on to our improved policy.

\stitle{Current {\kvcache} cache mechanism and policies. \,}
Regarding the mechanism:
existing systems cache {\kvcache}
primarily on GPU and CPU in a hierarchical manner.
The {\kvcache} is managed at block granularity (see \textsection{\ref{sec:bg}}):
when a {\kvcache} block is generated,
it is first cached on the GPU HBM.
If the GPU is unable to cache newly generated blocks due to out of memory,
the caching system adopts an eviction policy (described below)
to evict some blocks from the GPU to the CPU.
The evicted blocks are asynchronously swapped to CPU in a layer-by-layer manner
so the swap cost is negligible.
If the CPU is also out of memory,
the blocks are removed.
Before executing a new request,
if the request hits cached blocks,
the serving system directly reuses the cached blocks to avoid recomputation.
Note that if the hit is on CPU, the block is swapped to the GPU also in a layer-by-layer manner.

Regarding the policies: a {\kvcache} cache system mainly needs two policies:
how to determine cache hits and how to evict cached blocks.
For the first policy,
the design choices differ in their granularity of checking which blocks may contribute to the hits:
some systems check for matches across all requests~\cite{sglang,DBLP:journals/corr/abs-2404-12457}
while others only check for matches within requests from individual users~\cite{DBLP:journals/corr/abs-2403-14401, cachedattention}.
For the second policy,
current systems mainly adopt standard eviction policies like LRU or FIFO.
Though they adopt some extensions
like checking whether the evicted blocks are in the scheduling queue~\cite{cachedattention},
the overall rationale does not change.

\stitle{The drawback. \,}
Existing workload-agnostic policies
like LRU may miss the workload features we analyzed in \textsection{\ref{sec:analyze}},
and we argue that they are suboptimal.
Take Trace A as an example:
given the same amount of time since the last access (e.g., greater than 10 seconds),
10-turn text requests are more likely to have a subsequent turn than 1-turn text requests
(see {\fig{fig:motiv-next-turn-probability}}).
However, if many 1-turn requests are flooding into the system,
10-turn requests can be evicted by LRU
because the reuse time is longer.

\begin{table}[!t]
    \centering
    \small{
        \resizebox{1\linewidth}{!}{
            \ra{1.2}
\begin{tabular}{@{}p{0.45\linewidth}p{0.45\linewidth}@{}} \toprule
\textbf{Observations}                                                                                    & \textbf{Techniques}                                                      \\ \hline
The probability of {\kvcache} reuse follows exponential distributions given a workload 
({\fig{fig:motiv-next-turn-probability}} and {\fig{fig:motiv-reuse-time-probability-hist}}). 
& \ding{192} Workload-aware reuse-probability-distribution-based priority estimation.                   \\

Spatial locality ({\fig{fig:motiv-spatial}}--\ref{fig:motiv-spatial-workload}).                                                                                        
& \ding{193} Assign a higher priority to the head {\kvcache} blocks. \\

Ephemeral {\kvcache} lifespan ({\fig{fig:motiv-block-timespan}}).                                                          
& \ding{194} Remove frequency when calculating the priority,
as well as limiting the time span of probability calculation.                                     \\
\bottomrule
\end{tabular}
            
        }
    } \\[8pt]
    \begin{minipage}{1\linewidth}
        \caption{\small{\emph{
        A summary of the main techniques in our workload-aware cache eviction policy
        based on the findings in \textsection{\ref{sec:analyze}} 
        }}}
    \label{tab:insights}
    \end{minipage} \\[-5pt]
\end{table}

\stitle{Our solution: workload-aware reuse-probability-distribution-based {\kvcache} eviction policy. \,}
To this end, we propose a workload-aware cache eviction policy
that considers the reuse probability distributions
of each request category based on the characterizations
presented in \textsection{\ref{sec:motiv-temporal-spatial}}.
At a high level, at the eviction time, 
for each block, we calculate a priority---the likelihood of a {\kvcache} being reused
in the future based on the profiled reuse probability distribution 
of its request category (workload).
The priority further holistically considers the spatial locality as well as
the short lifespan of {\kvcache} blocks.
The blocks with the lowest priority are evicted first.
Table~\ref{tab:insights} summarizes our techniques
as well as their corresponding observations from \textsection{\ref{sec:analyze}}.

Our caching mechanisms follow 
existing systems~\cite{cachedattention, vllm, sglang, DBLP:journals/corr/abs-2403-14401}---including
asynchronous swapping and pipelined loading,
as these techniques have converged to a standard design.

\subsection{Workload-aware {\kvcache} cache eviction policy}
\label{sec:design-policy}

\nospacestitle{Existing workload-aware policy. \,}
Workload-aware cache eviction policy has been extensively studied in the literature~\cite{cherkasova1998improving,lruk,slru,lirs,arc,lhd,cacheus}.
A representative policy is the Greedy-Dual-Frequency-Size (GDFS) policy~\cite{cherkasova1998improving}, 
which evaluates the priority of each cached object (e.g., {\kvcache}) 
using the following workload-dependent features: 
$$
\text{Priority} = \text{Clock} + \frac{\text{Frequency} \times \text{Cost}}{\text{Size}}\,(\text{\uline{Existing GDFS}})
$$
The \texttt{Clock} captures the last access time of an object (similar to LRU),
the \texttt{Frequency} is the total access count of an object (similar to LFU),
the \texttt{Size} is the size of the cached object,
and \texttt{Cost} is the cost to bring the cached content back to the cache.
Objects with the lowest priority are evicted first.

While the above GDFS formulation can be retrofitted for calculating the priority of {\kvcache} blocks,
we found it does not fully consider the {\kvcache} usage in LLM serving
as well as the known probability distributions of requests in different categories.
For example, a frequently accessed block may not be reused in the future due to the short lifespan of {\kvcache} blocks,
causing a frequently accessed then died block wasting the {\kvcache} space.
Besides, a block with higher cost to bring to the cache
does not necessarily mean it is more likely to be reused in the future.
Instead, we need to bring the more likely to be reused blocks to the cache first.
While such workload information is not available in the setup of GDFS,
it is readily available in serving LLMs based on our characterization in the previous section.

\begin{figure}[!t]
    \begin{minipage}{.9\linewidth}
    \hspace{4mm}
    \includegraphics[width=1\textwidth, trim=0.25cm 11.8cm 24cm 0.25cm, clip]{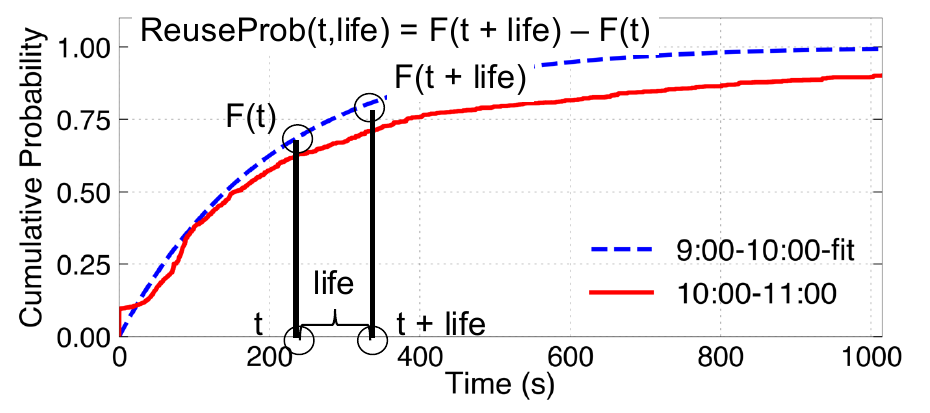} 
    \end{minipage} \\[0pt]
    \begin{minipage}{1\linewidth}
    \caption{\small{\textit{
        An illustration of how to calculate the reuse probability of a {\kvcache} block
        given its workload (request category). 
    }}}
    \label{fig:example-request-category}
    \end{minipage} \\[-15pt]
\end{figure}

\stitle{Our policy. \,}
We follow GDFS to calculate the priority of each {\kvcache} block and use
the priority to determine the eviction order,
but our priority is retrofitted with our characterized {\kvcache} reuse features:
$$
\text{Priority} = ({\text{ReuseProb}_w(t, \text{life})},{-\text{Offset}}) \,(\text{\uline{Workload-aware}})
$$
\noindent
Specifically, the priority is represented with a tuple containing two metrics
and we compare the tuple with lexicographical order.
The rationale behind the two metrics is: \\[-10pt]
\begin{itemize}[leftmargin=*,leftmargin=10pt,itemindent=0pt]
    \item \textbf{ReuseProb$_w$} (\ding{192}) calculates the reuse probability of a {\kvcache} block
    at a given time $t$, where $t$ means the time since its last access.
    For each workload (request category),
    the reuse probability is calculated using a two-step process.
    First, we sample a period of the recently accessed data in the background.
    Second, based on the observations in {\fig{fig:motiv-reuse-time-probability-hist}},
    we fit an exponential distribution to the sampled data,
    and look up the fitted curve to get the reuse probability.
    {\fig{fig:example-request-category}} shows an example of how to calculate the reuse probability.
    We first use the background sampled data from 9:00 a.m. to 10:00 a.m. to fit the cumulative distribution function (CDF) of the
    reuse probability of a given workload (blue line).

    Note that the fitted curve is close to the real data (red line).
    Afterward, we can look up the curve to get the reuse probability of a {\kvcache} block in the future.
    Note that the function further considers the expected lifespan (life) of the block,
    which is critical considering the short lifespan of {\kvcache} and we describe below.  \\[-10pt]

    \item \textbf{Offset} (\ding{193}) considers the spatial locality described in 
    \textsection{\ref{sec:motiv-temporal-spatial}}:
    the priority is reversely proportional to the prefix length of the {\kvcache} block in the entire request. \\[-10pt]
\end{itemize}

We don't consider frequency explicitly  (\ding{194})
because we found it can not provide useful information for cache policy
due to the short lifespan of {\kvcache} blocks (see {\fig{fig:motiv-block-timespan}}) even for frequently reused ones.
More importantly, we also need to regulate the calculated reuse probability
with the lifespan.
If not, the fitted reuse probability will return a large probability for a request
with long-tailed reuse time distribution for a long time.
This contradicts the short {\kvcache} lifespan observation
and leads to the inefficient utilization of limited cache space.

\begin{figure}[!t]
    \begin{minipage}{1\linewidth}
    \hspace{-1mm}
    \centering    
    \includegraphics[width=1\textwidth, trim=0.25cm 25.8cm 21cm 0.25cm, clip]{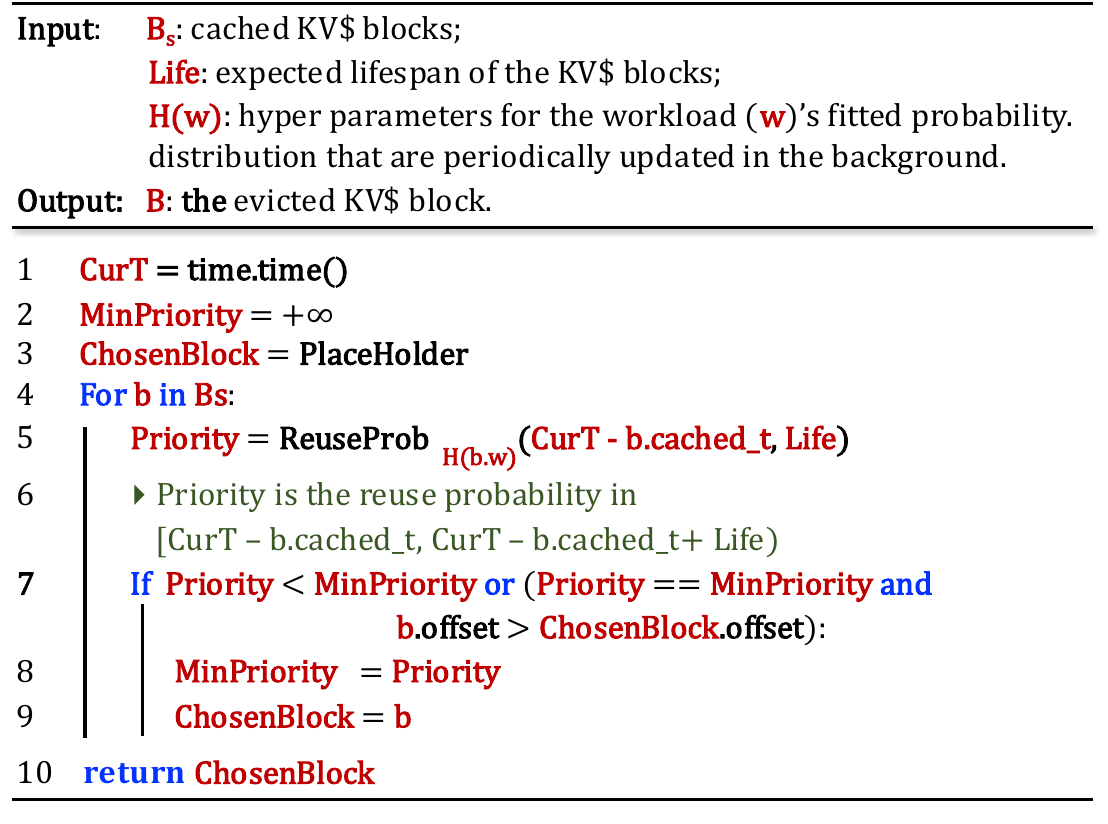} 
    \end{minipage} \\[4pt]
    \begin{minipage}{1\linewidth}
    \caption{\small{\textit{
        A simplified pseudocode of our workload-aware cache eviction algorithm.
    }}}
    \label{alg:naive-eviction}
    \end{minipage} \\[-5pt]
\end{figure}

\stitle{The detailed eviction algorithm. \,}
Algorithm~\ref{alg:naive-eviction} presents a simplified pseudocode of our workload-aware cache eviction algorithm.
Upon eviction,
we first iterate through all the cached blocks (line 4--6)
and calculate the reuse probability for each block using the workload ($b.w$).
The priority is then compared with the offset of the block
to update the block to be evicted (line 7--9).

\stitle{Performance optimization. \,}
One drawback of the above naive algorithm is that it
requires computing the reuse probabilities of all {\kvcache} blocks to identify the least reusable candidate.
To avoid excessive computation overhead,
we observe that all {\kvcache} blocks
within each workload are naturally ordered by their last accessed timestamps according to the exponential distribution,
so we can avoid computing the reuse probability when comparing blocks within the same workload.
Specifically, for each workload, we maintain
a priority queue of its {\kvcache} blocks ordered by last accessed time.
During eviction, instead of scanning all blocks, we first treat the least-recently-used block
of each workload as the candidate for eviction.
Afterward, we run the algorithm described in Algorithm~\ref{alg:naive-eviction}
to obtain the final evicted block.
This reduces the complexity from $O(N)$ to $O(W)$,
where $W$ is the number of workloads (typically in the range of tens).
As a result, the performance overhead of workload-aware policy is negligible to serving,
e.g., we observed a 79\,$\mu$s policy delay for each eviction,
with the aggregated overhead only 1.2\,\% of the scheduling overhead 
of a typical serving engine (vLLM~\cite{vllm}).

\subsection{Performance evaluation}
\label{sec:design-eval}

\nospacestitle{Testbed. \,}
We conducted our evaluations on a server with 8 NVIDIA A800-80GB GPUs.
The GPUs are connected via 400\,GBps bidirectional NVLink, 
allowing us to use multiple GPUs (e.g., 4) to serve a single large model
with tensor parallelism~\cite{narayanan2021efficient}. 
The GPUs are connected to the host via PCIe Gen4, 
offering a bidirectional bandwidth of up to 32\,GBps
between GPU HBM and the host memory.

\stitle{Implementations and baselines. \,}
Since the state-of-the-art {\kvcache} cache systems that
incorporate both CPU and GPU {\kvcache} like
CachedAttention~\cite{cachedattention} and Pensieve~\cite{DBLP:journals/corr/abs-2403-14401}
are not open-sourced,
we implement a CPU-GPU {\kvcache} cache on vLLM~\cite{vllm},
the state-of-the-art LLM serving system that adopts GPU {\kvcache} cache by default.
We have further integrated the policies and mechanisms from state-of-the-art
(see \textsection{\ref{sec:design-mechanism}}) to it.
We have calibrated that
our implementation has a similar performance to state-of-the-art like CachedAttention
on the same synthetic workload with a similar testbed.

Besides integrating a CPU {\kvcache} cache on vLLM,
we have also implemented the {\kvcache}-centric global scheduler described in Mooncake~\cite{mooncake},
because serving LLMs on multiple instances is common. 
Specifically, the global scheduler will record the global {\kvcache} cache usage
and schedule the request to the instance with cached {\kvcache} whenever possible. 
Note that our workload-aware caching policy only applies to a single serving instance,
extending it to a global layer is left as our future work. 

\stitle{Evaluated models and workloads. \,}
We use three representative open-source models for the performance evaluations: 
Qwen2-7B~\cite{qwen}, 
Llama2-13B~\cite{llama2} and Llama3-70B~\cite{DBLP:journals/corr/abs-2407-21783}.
The rationale for choosing these models and the differences between them have been 
discussed in the per-request {\kvcache} size analysis in \textsection{\ref{sec:analyze-capacity}}. 
For Qwen2-7B and Llama2-13B models, we use one GPU per serving instance. 
For Llama3-70B, we use 4 GPUs. 
We use the Trace A and B described in \textsection{\ref{sec:workload-data}} as our evaluated workloads. 
Since we don't have sufficient GPUs to support the real workload,
we scale the traces to fit the processing capability of our testbed
using the scaling method that preserves the temporal pattern of the trace~\cite{DBLP:conf/eurosys/SajalHZU021}. 

\stitle{Evaluation methodology on anonymous traces. \,}
Even though the traces are anonymous,
we can evaluate the system serving performance
by constructing inputs that accurately reproduce the original trace's cache hit patterns.
Specifically,
for each anonymous block without assigned tokens,
we randomly generate a token list, ensuring each generated list's hash differs from all other token lists.
Moreover, to match the original input's iteration count,
we constrain the model using the trace's output length.
Finally, we replace the model-generated output with the constructed tokens
to preserve the cache patterns of multi-turn requests.

\begin{figure}[!t]
    \centering
    \includegraphics[width=1.0\linewidth,center]{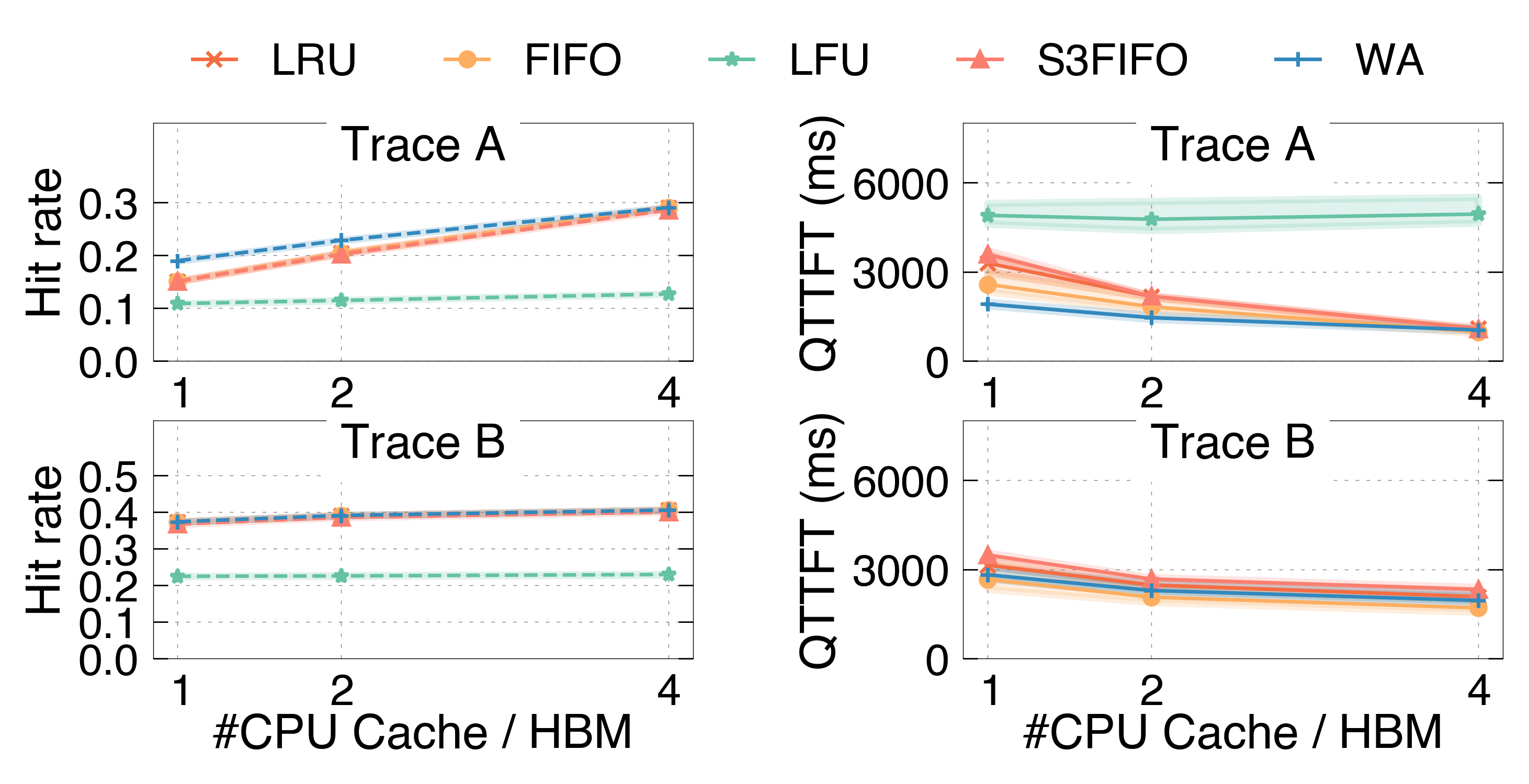}  \\[1pt]
    \begin{minipage}{1\linewidth}
    \caption{\small{\emph{      
        An analysis of the cache hit ratio and QTTFT with respect to the 
        CPU cache provisioned on Qwen2-7B.   
    }}}
    \label{fig:eval-qwen7b-cpu-cache}
    \end{minipage} \\[-5pt]
\end{figure} 

\begin{figure}[!t]
    \centering
    \includegraphics[width=1.0\linewidth,center]{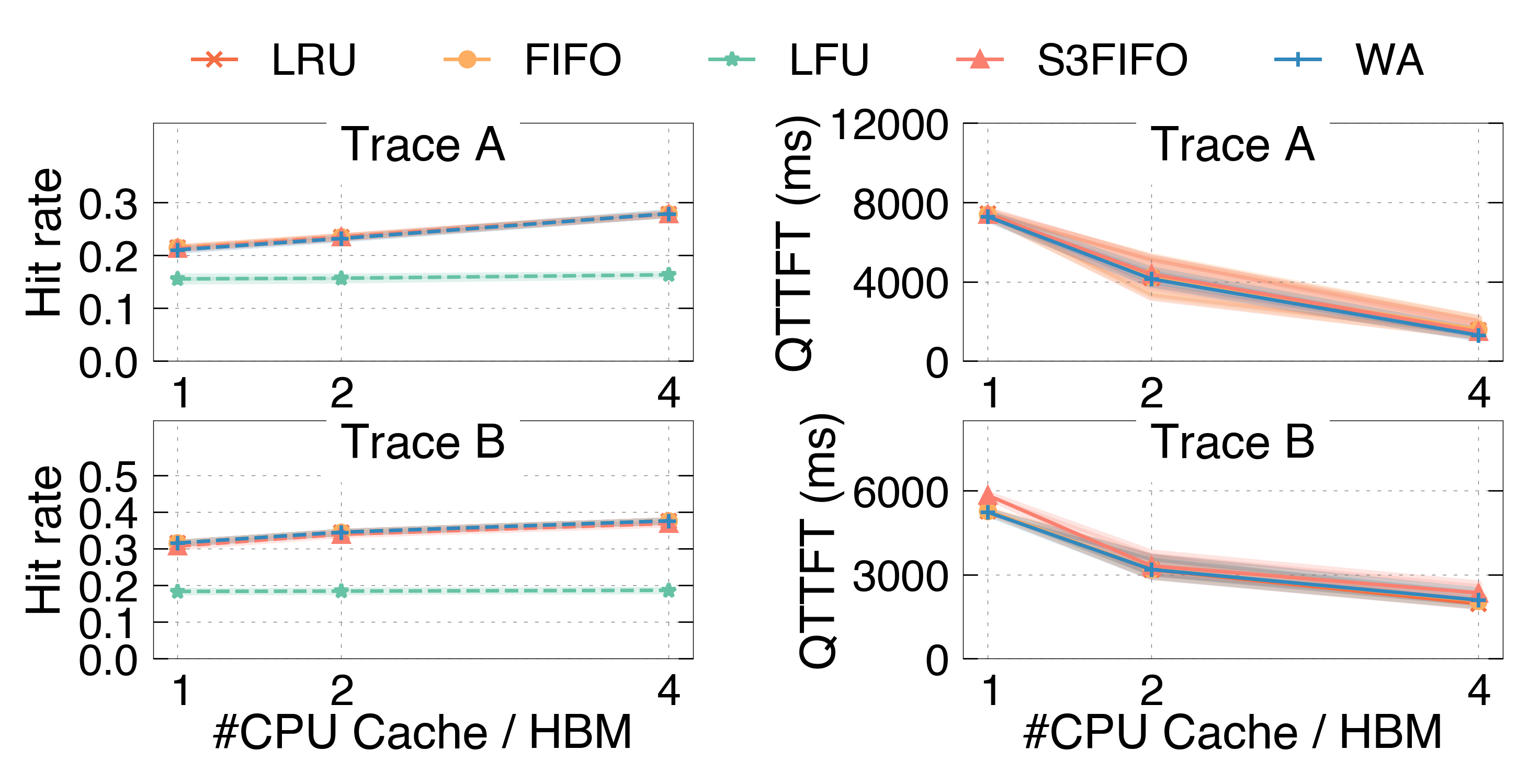}  \\[1pt]
    \begin{minipage}{1\linewidth}
    \caption{\small{\emph{        
        An analysis of the cache hit ratio and QTTFT with respect to the 
        CPU cache provisioned on Llama2-13B.           
    }}}
    \label{fig:eval-llama13b-cpu-cache}
    \end{minipage} \\[-5pt]
\end{figure} 

\begin{figure}[!t]
    \centering
    \includegraphics[width=1.0\linewidth,center]{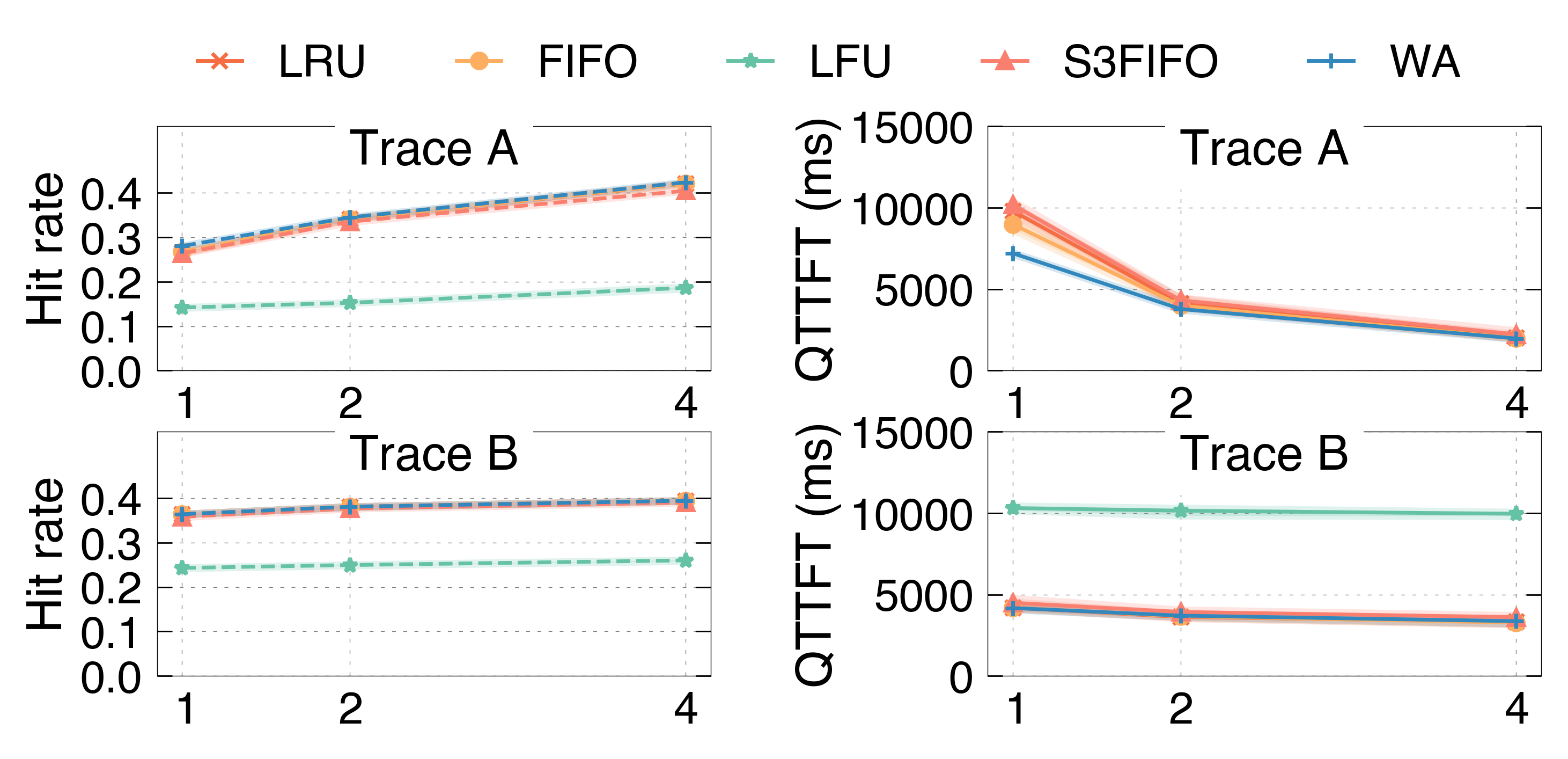}  \\[1pt]
    \begin{minipage}{1\linewidth}
    \caption{\small{\emph{     
        An analysis of the cache hit ratio and QTTFT with respect to the 
        CPU cache provisioned on Llama3-70B.            
    }}}
    \label{fig:eval-llama70b-cpu-cache}
    \end{minipage} \\[-5pt]
\end{figure} 

\begin{figure}[!t]
    \centering
    \includegraphics[width=1.0\linewidth,center]{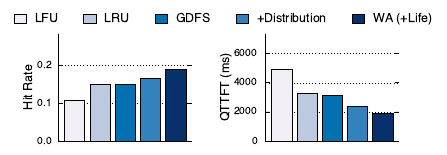}  \\[3pt]
    \begin{minipage}{1\linewidth}
    \caption{\small{\emph{     
        An ablation study on Qwen2-7B with \#CPU Cache / HBM is 1.
    }}}
    \label{fig:cache-policy-abl}
    \end{minipage} \\[-5pt]
\end{figure}

\stitle{Improved cache hit. \,}
{\fig{fig:eval-qwen7b-cpu-cache}}--{\ref{fig:eval-llama70b-cpu-cache}} 
show the cache hit ratio and serving performance 
for standard cache policies and our workload-aware policy (WA). 
We compare our policy with the standard policies like 
least-recently-used (LRU), first-in-first-out (FIFO),
least-frequently-used (LFU) and recent optimized 
FIFO method (S3-FIFO)~\cite{10.1145/3600006.3613147}. 
We compare with the vanilla GDFS in {\fig{fig:cache-policy-abl}}. 

Since the performance depends on cache capacity,
we report the hit rates (as well as performance) by varying the provisioned {\kvcache} cache capacity.
The capacity is normalized to the GPU HBM size for {\kvcache},
i.e., the \#CPU Cache / HBM in the figures means
 the CPU memory used to store the {\kvcache} normalized to the GPU HBM.
 For example, a \#CPU Cache / HBM of 1 means the CPU memory used to store the {\kvcache}
 is equal to the GPU HBM size.
We draw three observations from the results.

First, our policy {\policy} achieves 8.1--23.9\,\% higher hit rate compared to other baselines,
and it has a 1.5--3.9\,\% higher hit rate compared to the best of other baselines.
Our policy is more effective thanks to its knowledge of the workload reuse patterns.
Second, the effectiveness of {\policy} decreases when the cache capacity increases,
this is as expected because a larger cache can tolerate a less effective policy.
Finally, WA is more effective on Trace A than Trace B, this is due to two reasons. 
(1) Trace B has little workload information (> 99\,\% of its requests are 1-turn requests).
Thus, our fitted exponential distribution-based policy essentially falls back
to a normal LRU. (2) Trace B requires a relatively small {\kvcache} capacity,
so GPU HBM is already close to optimal except for LFU.
The poor hit rate of LFU comes from the fact that the most frequently accessed blocks may die early
thus polluting the {\kvcache} cache, as we have extensively discussed before.

\stitle{Improved performance. \,}
The improved cache hit ratio leads to a significant performance improvement 
in the serving performance. 
{\fig{fig:eval-qwen7b-cpu-cache}}--{\ref{fig:eval-llama70b-cpu-cache}} also reports 
the queued TTFT (QTTFT) when serving different traces,
i.e., the TTFT plus the time when waiting for GPU to process the request. 
We can see that {\policy} achieves 28.3--41.9\% QTTFT reduction thanks to the improved 
{\kvcache} cache hit rate. 

\stitle{Ablation study. \,}
{\fig{fig:cache-policy-abl}} conducts an ablation study on the Qwen2-7B model on Trace A
with different policies. Other models share a similar trend.
First, we can see that using a distribution-based approach improves the
hit rate by 1\,\% (with technique \ding{192} in Table~\ref{tab:insights}).
Considering the short lifespan of {\kvcache} with the life regulator further
increases the hit rate by 2.4\,\% (\ding{194}).
Note that we omit describing the locality optimization \ding{193} as it is applied
to all the policies.

\stitle{Discussion: fairness issue. \,}
Like all existing systems (and their policies),
a malicious client can monopolize prefix cache by sending a large volume of requests with long identical prefixes,
thus causing fairness issues.
Ensuring fairness is an orthogonal topic to cache policy
and can be addressed via a co-design of LLM serving strategies like FairRide~\cite{DBLP:conf/nsdi/PuLZGS16,DBLP:conf/osdi/0007CLZ0ZGS24}.

\section{Related Work}
\label{sec:related}

\nospacestitle{Reuse {\kvcache} across requests via {\kvcache} cache. \,}
Reusing {\kvcache} for accelerating LLM serving
has been widely studied~\cite{vllm,chunkattention,cachedattention,promptcache,sglang,cacheblend} and used in commercial LLM serving~\cite{openaiapi,genimiapi,claudeapi,mooncake}.
Currently, production systems only treat cache hits by prefix matches~\cite{chunkattention,sglang,cachedattention,openaiapi,genimiapi,claudeapi,mooncake}, 
because it preserves the original algorithm of model inference with no accuracy loss.
A number of studies~\cite{promptcache,cacheblend,hu2024epic} have studied 
methods for non-prefix {\kvcache} cache. 
We can revisit our characterization once they have matured 
and been deployed in production.

\stitle{Other {\kvcache}-related optimizations. \,}
Emerging works\cite{streamingllm,h2o,infinigen,pyramidkv} propose runtime {\kvcache}
compression and deletion methods to reduce {\kvcache} size.
StreamingLLM~\cite{streamingllm} keeps
only a finite attention window for recent tokens combined with a few initial tokens.
H2O~\cite{h2o} only involves tokens that contribute most to the attention score for inference.
InfiniGen~\cite{infinigen} speculatively selects tokens critical for attention scores and drops others.
KVQuant~\cite{KVQuant} enables 3-bit KV cache quantization with minimal perplexity degradation 
for large language models, achieving fast million-length context inference.
These methods require fewer {\kvcache} per-request,
albeit with accuracy degradation.
Our study is compatible with these works:
if an uncompressed/full {\kvcache} is reusable,
a compressed or deleted version would also be reusable.

\stitle{Optimizing caching policies. \,}
Optimizing caching has long been studied in the literature,
in general-purpose caching policies~\cite{lruk,slru,twoq,eelru,lrfu,lirs,arc,mq,car,clockpro,DBLP:journals/tos/EinzigerEFM22,lhd,cacheus,sieve,cherkasova1998improving}
or specific domains~\cite{yang2020twemcache,berg2020cachelib,icebreaker,fasscache}.
We continue this line of research,
and leverage our characterized {\kvcache} reuse properties
for optimizing {\kvcache} cache policies.

\stitle{Optimizing LLM serving. \,}
We continue the line of research in optimizing the performance of LLM serving systems~\cite{DBLP:conf/osdi/ZhongLCHZL0024,DBLP:journals/corr/abs-2404-09526,DBLP:conf/sosp/KwonLZ0ZY0ZS23,alpaserve,DBLP:conf/osdi/YuJKKC22,DBLP:conf/isca/PatelCZSGMB24,cachedattention,DBLP:journals/corr/abs-2412-17246},
with a particular focus on characterizing serving workloads
for {\kvcache} cache.
Our work is orthogonal to optimizations other than {\kvcache} cache.

\section{Conclusion}
\noindent
We present the first systematic and in-depth characterization
of production serving workloads for {\kvcache} cache systems.
Our study reveals several important findings
that were not captured by synthetic workloads used in previous studies.
Based on these findings,
we further propose a workload-aware cache eviction policy
that improves upon existing workload-agnostic policies like LRU and LFU.
We believe our work, as a starting point,
can benefit both the research and industry fields
in improving current and future LLM serving systems with a workload-driven approach.

\section*{Acknowledgment}
\noindent
We would like to thank ATC reviewers and our shepherd Lei Zhang for
their insightful feedback.
We sincerely thank Guanhu Wang for helping us with the collection of the trace data.
We also thank Xiating Xie for proofreading our paper. 
We thank Xiangfan Wu for helping us correct the presentation about hashing the token IDs.
Furthermore, we thank Zhijun Yang for refining the presentation of this paper.
This work was supported in part by
the National Natural Science Foundation of China (No. 62202291, 62272291, 62132014) and
the Fundamental Research Funds for the Central Universities. 
This work was also supported by
a research grant from Alibaba Group through the Alibaba Innovative Research Program. 

\balance

\small{
\bibliographystyle{acm}
\bibliography{kvcache}
}

\appendix

\clearpage
\section{Appendix}

\subsection{Trace results across different days}
\label{analysis:reuse-evolution}

\begin{figure*}[!t]
    \centering
    \includegraphics[width=1.0\linewidth,center]{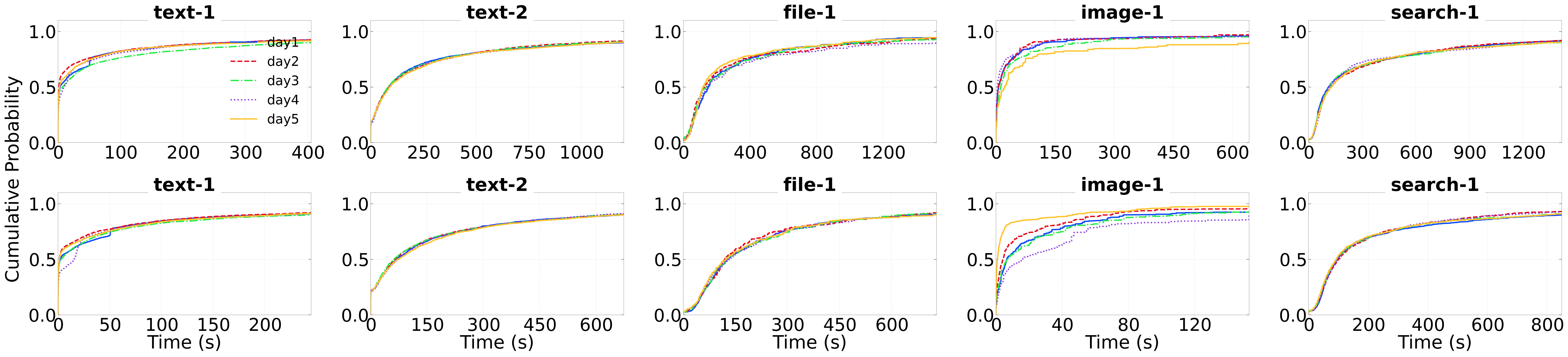} \\[5pt]
    \begin{minipage}{1\linewidth}
    \caption{\small{\emph{
        The evolution of cumulative distribution function of the reuse probability in 5 workdays.
        The first row: distributions from 9.00 a.m. to 10.00 a.m.
        The second row: distributions from 10.00 a.m. to 11.00 a.m.
    }}}
    \label{fig:motiv-7days}
    \end{minipage} \\[-5pt]
\end{figure*}

\noindent
\fig{fig:motiv-7days} shows the evolution of cumulative distribution (CDF) function of 
the reuse probability for each request category in future time window over 5 continuous workdays.
We can observe that the reuse probability at 9:00 a.m. to 10:00 a.m.
similar between workdays. Therefore, we can analyze the reuse log
of last day offline to predict the reuse probability of each category 
in the next day. However, it is noted that the reuse probability of 
"image" type is harder to predict than other categories, 
we infer that this is because image types have coarser reusability granularity 
compared to text types, and there exists greater variability among different users.

\end{document}